\documentclass[prl,twocolumn,superscriptaddress,nobibnotes,letterpaper]{revtex4-1}

\usepackage[pdftex]{graphicx}
\usepackage[pdftex]{epsfig}
\usepackage{amsmath}
\usepackage{amssymb}
\usepackage{amsfonts}
\usepackage{color}
\usepackage{wrapfig}
\usepackage{eucal}
\usepackage{hhline}
\usepackage{threeparttable}
\usepackage{supertabular}
\usepackage{multirow}
\usepackage{tabularx}

\usepackage[colorlinks=true, allcolors=blue]{hyperref}

\newcommand{\opd}[2]{\mbox{$\hat{#1}_\text{#2}^{\dagger}$}}
\newcommand{\op}[2]{\mbox{$\hat{#1}_\text{#2}$}}

\newcommand{\ket}[2]{\mbox{$\rvert{#1}\rangle_\text{#2}$}}

\def\be{\begin{equation}}
\def\ee{\end{equation}}
\def\bea{\begin{eqnarray}}
\def\eea{\end{eqnarray}}

\newcolumntype{Y}{>{\centering\arraybackslash}X}

\begin{document}

\pagenumbering{arabic}

\title{Remote quantum entanglement between two micromechanical oscillators}\thanks{This work was published in Nature \textbf{556}, 473--477 (2018).}

\author{Ralf Riedinger}\thanks{These authors contributed equally to this work.}
\affiliation{Vienna Center for Quantum Science and Technology (VCQ), Faculty of Physics, University of Vienna, A-1090 Vienna, Austria}
\author{Andreas Wallucks}\thanks{These authors contributed equally to this work.}
\affiliation{Kavli Institute of Nanoscience, Delft University of Technology, 2628CJ Delft, The Netherlands}
\author{Igor Marinkovi\'{c}}\thanks{These authors contributed equally to this work.}
\affiliation{Kavli Institute of Nanoscience, Delft University of Technology, 2628CJ Delft, The Netherlands}
\author{Clemens L\"oschnauer}
\affiliation{Vienna Center for Quantum Science and Technology (VCQ), Faculty of Physics, University of Vienna, A-1090 Vienna, Austria}
\author{Markus Aspelmeyer}
\affiliation{Vienna Center for Quantum Science and Technology (VCQ), Faculty of Physics, University of Vienna, A-1090 Vienna, Austria}
\author{Sungkun Hong}
\email{sungkun.hong@univie.ac.at}
\affiliation{Vienna Center for Quantum Science and Technology (VCQ), Faculty of Physics, University of Vienna, A-1090 Vienna, Austria}
\author{Simon Gr\"oblacher}
\email{s.groeblacher@tudelft.nl}
\affiliation{Kavli Institute of Nanoscience, Delft University of Technology, 2628CJ Delft, The Netherlands}

\begin{abstract}
Entanglement, an essential feature of quantum theory that allows for inseparable quantum correlations to be shared between distant parties, is a crucial resource for quantum networks~\cite{Kimble2008}. Of particular importance is the ability to distribute entanglement between remote objects that can also serve as quantum memories. This has been previously realized using systems such as warm~\cite{Jensen2011,Reim2011} and cold atomic vapours~\cite{Chou2005,Matsukevich2006}, individual atoms~\cite{Ritter2012} and ions~\cite{Moehring2007,Jost2009}, and defects in solid-state systems~\cite{Usmani2012,Saglamyurek2015,Hensen2015}. Practical communication applications require a combination of several advantageous features, such as a particular operating wavelength, high bandwidth and long memory lifetimes. Here we introduce a purely micromachined solid-state platform in the form of chip-based optomechanical resonators made of nanostructured silicon beams. We create and demonstrate entanglement between two micromechanical oscillators across two chips that are separated by 20 centimetres. The entangled quantum state is distributed by an optical field at a designed wavelength near 1550 nanometres. Therefore, our system can be directly incorporated in a realistic fibre-optic quantum network operating in the conventional optical telecommunication band. Our results are an important step towards the development of large-area quantum networks based on silicon photonics.
\end{abstract}

\maketitle

In recent years, nanofabricated mechanical oscillators have emerged as a promising platform for quantum information processing. The field of opto- and electromechanics has seen great progress,
including ground-state cooling~\cite{Teufel2011b,Chan2011}, quantum interfaces to optical or microwave modes~\cite{Palomaki2013,Riedinger2016}, mechanical squeezing~\cite{Wollman2015,Pirkkalainen2015,Lecocq2015} and single-phonon manipulation~\cite{OConnell2010,Chu2017,Hong2017,Reed2017}. Demonstrations of distributed mechanical entanglement, however, have so far been limited to intrinsic material resonances~\cite{Lee2011} and the motion of trapped ions~\cite{Jost2009}. Entanglement of
engineered (opto-)mechanical resonances, on the other hand, would provide a route towards scalable quantum networks. The freedom of designing and choosing optical resonances would allow operation in the entire frequency range of the technologically important C-, S- and
L-bands of fibre-optic telecommunications. Together with dense wavelength-division multiplexing (on the ITU-T grid), this could enable quantum nodes separated by long distances (about 100~km) that can communicate at large bandwidths. State-of-the-art engineered mechanical elements have energy lifetimes that typically range between micro-~\cite{Riedinger2016} and milliseconds~\cite{Meenehan2015}, which would allow entanglement distribution on a regional level~\cite{Razavi2009}. In addition, these entangled mechanical systems could be interfaced with microwaves~\cite{Bochmann2013}, opening up the possibility of integrating superconducting quantum processors in the local nodes of the network.

Here we report on the observation of distributed entanglement between two nanomechanical resonators, mediated by telecommunication-wavelength photons. We use the DLCZ protocol~\cite{Duan2001}, which was experimentally pioneered with ensembles of cold atoms~\cite{Chou2005}. The entanglement is generated probabilistically through the conditional preparation of a single phonon, heralded by the detection of a signal photon that could originate from either of two identical optomechanical oscillators. Fabrication imperfections have previously limited the use of artificial structures, requiring external tuning mechanisms to render such systems indistinguishable. Here we demonstrate not only that obtaining sufficiently identical devices is in fact possible through nanofabrication, but also that our method could in principle be applied to more than two systems.

\begin{figure*}[ht]
	\begin{center}
		\includegraphics[width=1.9\columnwidth]{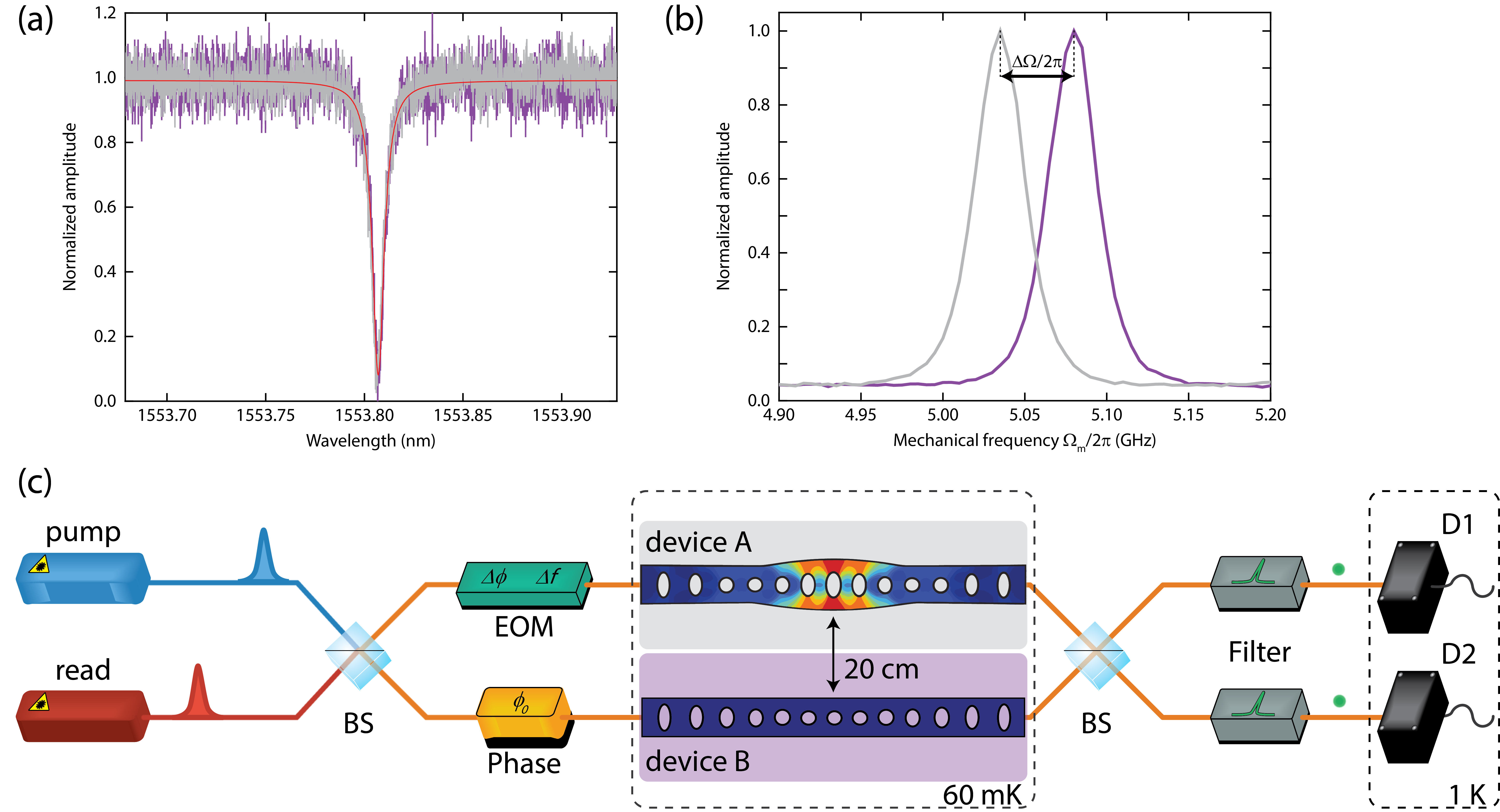}
		\caption{\textbf{Devices and experimental setup.} \textbf{a}, Optical resonances of device A (grey) and device B (magenta). The Lorentzian fit result (red line) yields a quality factor of $Q=2.2\times10^5$ for each cavity. \textbf{b}, Mechanical resonances of device A (grey) and device B (magenta). The normalized mechanical resonances are measured through the optomechanical sideband scattering rates. The linewidth is limited by the bandwidth of the optical pulses and filters. The frequencies of the devices differ by $\Delta\Omega_\textrm{m}/2\pi=45$~MHz, which could result in distinguishable photons, potentially reducing the entanglement in the system. We compensate for this shift by tuning	the optical pump fields accordingly through serrodyning, erasing any information that could lead to a separable state. \textbf{c}, Experimental setup. We create optical pulses using two lasers, which are detuned to the Stokes (pump) and anti-Stokes (read) transition of the optomechanical cavities. The lasers are then combined on a 50/50 beam splitter (BS), which forms an interferometer with a second combining beam splitter. Each arm of the interferometer contains one of the mechanical oscillators, cooled to its ground state using a dilution refrigerator (central dashed rectangle). The phase of the interferometer, $\phi_0$, is stabilized using a fibre stretcher (labelled `phase'), while the phase difference between the pulses, $\Delta\phi$, is controlled	using an electro-optic modulator (EOM). The same EOM is also used for serrodyning. Optical filters in front of two superconducting single-photon detectors (D1, D2) ensure that only photons scattered onto the cavity resonance are detected, whereas the original laser pulses are completely suppressed. The mechanical devices are physically separated by 20~cm and their optical separation is around 70~m.}
		\label{fig:1}
	\end{center} 
\end{figure*}

The mechanical oscillators that we use in our experiment are nano-structured silicon beams with co-localized mechanical and optical resonances. Radiation pressure forces and the photoelastic effect couple the optical and mechanical modes with a rate $g_0$, causing the optical frequency to shift under the displacement of the mechanical oscillator~\cite{ChanPhD}. This effect can be used to selectively address Stokes and anti-Stokes transitions by driving the optical resonance with detuned laser beams, resulting in a linear optomechanical interaction. As was recently shown, this technique can be used to create non-classical mechanical and optomechanical states at the single-quantum level for individual devices by using photon counting and post-selection~\cite{Riedinger2016,Hong2017}.

To apply the DLCZ scheme to the entanglement of two separate optomechanical crystals, a critical requirement is that the photons emitted from the optomechanical cavities must be indistinguishable. This can be achieved by creating a pair of nanobeams with identical optical
and mechanical resonances. Until now, however, fabrication variations have inhibited the deterministic generation of identical devices and the design of current oscillators does not include any tuning capabilities. Considering the optical mode alone, typical fabrication runs result in a spread of the resonance frequency of about 2~nm around the centre wavelength. Therefore, finding a pair of matching optical resonances on two chips close to a target frequency currently relies on fabricating a large enough set, in which the probability of obtaining an identical pair is sufficiently high. In fact, this is achievable with a few hundred
devices per chip (see Supplementary Information for details). In addition, a small mismatch in the mechanical frequencies, which is typically around 1\%, can readily be compensated by appropriate manipulation of the optical pulse frequencies in the experiment.

For the experiments presented here, we chose a pair of devices with optical resonances at wavelength $\lambda=1553.8$~nm (optical quality factor $Q=2.2\times10^5$ and $g_0/2\pi=550$~kHz and $790$~kHz  for devices A and B, respectively; see Fig.~\ref{fig:1}). For these structures, the mechanical resonance frequencies are centred around $\Omega_\textrm{m}/2\pi\approx5.1$~GHz and have a difference of $\Delta\Omega_\textrm{m}/2\pi=45$~MHz. The two chips are mounted 20~cm apart in a dilution refrigerator. Although we use a single cryostat, there is in principle no fundamental or technical reason for keeping the devices in a common cold environment. For our setup, if the telecommunication fibres linking the two devices were to be unwrapped, our setup would already allow us to bridge a separation of about 70~m between the two chips without further modification.

\begin{figure}[t]
	\begin{center}
		\includegraphics[width=0.9\columnwidth]{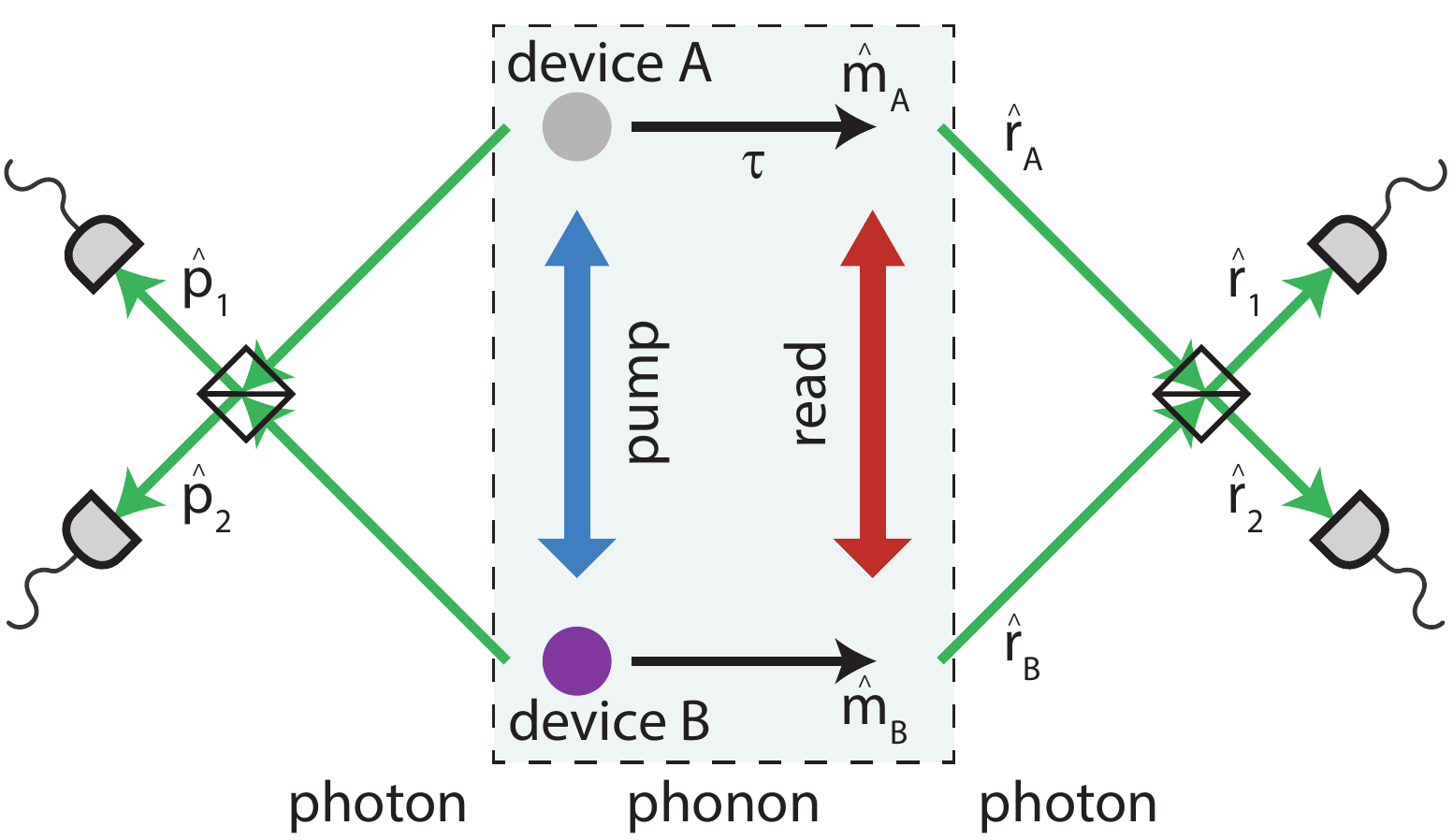}
		\caption{\textbf{Creation and detection of entanglement between two remote mechanical oscillators.} A pump pulse detuned to the Stokes sideband of two identical optomechanical resonators is sent into an interferometer, creating a single excitation in either device A or B. This process emits a photon on resonance with one of the cavities, and the two possible paths are superimposed using a beam splitter (black square) when exiting the interferometer (left). Detection of this photon in one of the single-photon detectors projects the two mechanical systems into an entangled state, in which neither device can be described separately. To verify this non-separable state, an optical read pulse tuned to the anti-Stokes sideband is sent into the interferometer with a delay of $\tau$, de-exciting the mechanical systems and emitting another on-resonance photon into modes $r_i$ ($i=\textrm{A,B}$) with operators \op{r}{i}. The two optical paths are again superimposed on the same beam splitter (right), and the photon is detected, allowing us to measure various second-order correlation functions, which are used to test an entanglement witness. The operators \op{p}{j} and \op{r}{j}, with $j=1,2$, denote the optical modes created from the pump and the read pulses, respectively, after recombination on the beam splitter and \op{m}{i} ($i=\textrm{A,B}$) are the operators of the mechanical modes. We note that in our experiment, the detectors used for the pump and read photons are identical (see Fig.~\ref{fig:1}).}
		\label{fig:2}
	\end{center} 
\end{figure}

The protocol~\cite{Duan2001} for the creation and verification of the remote mechanical entanglement consists of three steps (for a schematic, see Fig.~\ref{fig:2}). First, the two mechanical resonators are cryogenically cooled, and thus initialized close to their quantum ground states~\cite{Meenehan2015,Riedinger2016,Hong2017} (see Supplementary Information). Second, a weak `pump' pulse tuned to the upper mechanical sideband (at frequency $\omega_\textrm{pump}=2\pi c/\lambda+\Omega_\textrm{m}$, where $c$ is the speed of light), is sent into a phase-stabilized interferometer (with a fixed phase difference $\phi_0$, see Fig.~\ref{fig:1} and Supplementary Information) with one device in each arm. This drives the Stokes process--that is, the scattering of a pump photon into the cavity resonance while simultaneously creating a phonon~\cite{Riedinger2016}. The presence of a single phonon is heralded by the detection of a scattered Stokes photon in one of our superconducting nanowire single-photon detectors. The two optical paths of the interferometer are overlapped on a beam splitter, and a variable optical attenuator is set on one of the arms so that a scattered photon from either device is equally likely to reach either detector. The heralding detection event therefore contains no information about which device the scattering took place in and thus where the phonon was created. The energy of the pulse is tuned to ensure that the scattering probability $p_\textrm{pump}\approx 0.7\%$ is low, making the likelihood of simultaneously creating phonons in both devices negligible. The heralding measurement therefore projects the mechanical state into a superposition of a single-excitation state in device A $(\ket{A}{}=\ket{1}{A} \ket{0}{B})$ or device B $(\ket{B}{}=\ket{0}{A} \ket{1}{B})$, with the other device remaining in the ground state. The joint state of the two mechanical systems
\begin{eqnarray}
\ket{\Psi} = \frac{1}{\sqrt{2}} \left( \ket{1}{A} \ket{0}{B} \pm e^{i\theta_\mathrm{m}(0)} \ket{0}{A} \ket{1}{B} \right)\label{eq:MechPsi}
\end{eqnarray}
is therefore entangled, where $\theta_\mathrm{m}(0) = \phi_0$ is the phase with which the mechanical state is initialized at delay  $\tau=0$. This phase is determined from the relative phase difference that the pump beam acquires in the two interferometer arms~\cite{Chou2005}, which we can choose using our interferometer lock. However, because the two mechanical frequencies differ by $\Delta \Omega_\mathrm{m}$, the phase of the entangled state will continue to evolve as $\theta_\mathrm{m}(\tau) = \phi_0 + \Delta \Omega_\mathrm{m}\tau$. The sign in equation~\eqref{eq:MechPsi} reflects which detector is used for heralding, with $+$ ($-$) corresponding to the positive (negative) detector, as defined by the sign convention of the interferometer phase $\phi_0$.

In the third step of our protocol, we experimentally verify the entanglement between the two mechanical oscillators. To achieve this, we map the mechanical state onto an optical field using a `read' pulse after a variable delay $\tau$. This relatively strong pulse is tuned to the lower mechanical sideband of the optical resonance ($\omega_\textrm{read}=2\pi c/\lambda-\Omega_m$). At this detuning, the field drives the anti-Stokes transition--that is, a pump photon is scattered onto the cavity resonance while annihilating a phonon~\cite{Riedinger2016}. Ideally, this state transfer will convert $\ket{\Psi}{}$ into
\begin{eqnarray}
\ket{\Phi} = \frac{1}{\sqrt{2}} \left( \ket{1}r_\mathrm{\scriptscriptstyle A} \ket{0}r_\mathrm{\scriptscriptstyle B} \pm e^{i(\theta_\mathrm{r}+\theta_\mathrm{m}(\tau))} \ket{0}r_\mathrm{\scriptscriptstyle A} \ket{1}r_\mathrm{\scriptscriptstyle B} \right),
\label{eq:state}
\end{eqnarray}
where $\mathrm{r_A}$ and $\mathrm{r_B}$ are the optical modes in the two interferometer arms. The state of the optical field now contains the mechanical phase as well as the phase difference $\theta_\mathrm{r}$ acquired by the read pulse. We can add an additional phase offset $\Delta \phi$ to the read pulse in one of the interferometer arms so that $\theta_\mathrm{r} = \phi_0 + \Delta \phi$ by using an electro-optic phase modulator, as shown in Fig.~\ref{fig:1}. Sweeping $\Delta \phi$ allows us to probe the relative phase $\theta_\mathrm{m}(\tau)$ between the superpositions $\ket{A}{}$ and $\ket{B}{}$ of the mechanical state for fixed delays $\tau$. To avoid substantial absorption heating creating thermal excitations in the oscillators, we limit the energy of the read pulse to a state-swap fidelity of about 3.4\%, reducing the number of added incoherent phonons to about 0.07 at a delay of $\tau=123$~ns (see Supplementary Information).

So far we have neglected the consequence of slightly differing mechanical resonance frequencies for our heralding scheme. To compensate for the resulting frequency offset in the scattered (anti-) Stokes photons and to erase any available `which device' information, we shift the frequency of the laser pulses by means of serrodyning (see Supplementary Information). Specifically, we use the electro-optic phase modulator, which controls the phase offset $\Delta\phi$, to also shift the frequency of the pump (read) pulses to device A by $+\Delta\Omega_\textrm{m}$ ($-\Delta\Omega_\textrm{m}$). The frequency differences of the pulses in the two opposing paths cancel out their mechanical frequency differences exactly, ensuring
that the scattered photons at the output of the interferometer are indistinguishable.

To confirm that the measured state is indeed entangled, we need to distinguish it from all possible separable states, that is, the set of all states for which systems A and B can be described independently. A specifically tailored measure that can be used to verify this non-separability of the state is called an `entanglement witness'. Here we use a witness that is designed for optomechanical systems~\cite{Borkje2011}. In contrast to other path-entanglement witnesses based on partial state tomography, such as concurrence, this approach replaces measurements of third-order coherences, $g^{(3)}$, by expressing them as second-order coherences, $g^{(2)}$, assuming linear interactions between Gaussian states. This greatly simplifies the requirements and reduces the measurement times for our experiments. Because the coherences refer to the unconditional states, the nonlinear detection and state projection do not contradict these assumptions. The above assumptions are satisfied for our system because the initial mechanical states of our devices are in fact thermal states close to the corresponding quantum ground states (step 1 of our protocol; see Supplementary Information) and we use linearized optomechanical interactions (described in steps 2 and 3)~\cite{Wieczorek2015}. The upper bound for this witness of mechanical entanglement is given by~\cite{Borkje2011} (see Supplementary Information).
\begin{equation}
R_\textrm{m} (\theta, j)= 4\cdot \frac{g^{(2)}_{r_1,p_j}(\theta)+g^{(2)}_{r_2,p_j}(\theta)-1}{(g^{(2)}_{r_1,p_j}(\theta)-g^{(2)}_{r_2,p_j}(\theta))^2},
\label{eq:witness}
\end{equation}
in a symmetric setup. In equation~\eqref{eq:witness}, $\theta = \theta_\mathrm{r} + \theta_\mathrm{m}$, $j=1,2$ denotes the heralding detectors and $g^{(2)}_{r_i,p_j}=\langle \opd{r}{i} \opd{p}{j} \op{r}{i} \op{p}{j}\rangle/\langle \opd{r}{i} \op{r}{i} \rangle \langle  \opd{p}{j} \op{p}{j}\rangle$ is the second-order coherence between the photons scattered by the pump pulse (with $\opd{p}{j}$ and $\op{p}{j}$ the creation and annihilation operators, respectively, of the mode going to detector $j$) and the converted phonons from the read pulse (with $\opd{r}{j}$ and $\op{r}{j}$ the creation and annihilation operators, respectively, of the mode going to detector $j$). For all separable states of the mechanical oscillators A and B, the witness yields $R_\textrm{m}(\theta, j)\geq 1$ for any $\theta$ and $j$. Hence, if there exists a $\theta$ and $j$ for which $R_\textrm{m}(\theta, j)<1$, the mechanical systems must be entangled.

\begin{figure}[t]
	\begin{center}
		\includegraphics[width=0.9\columnwidth]{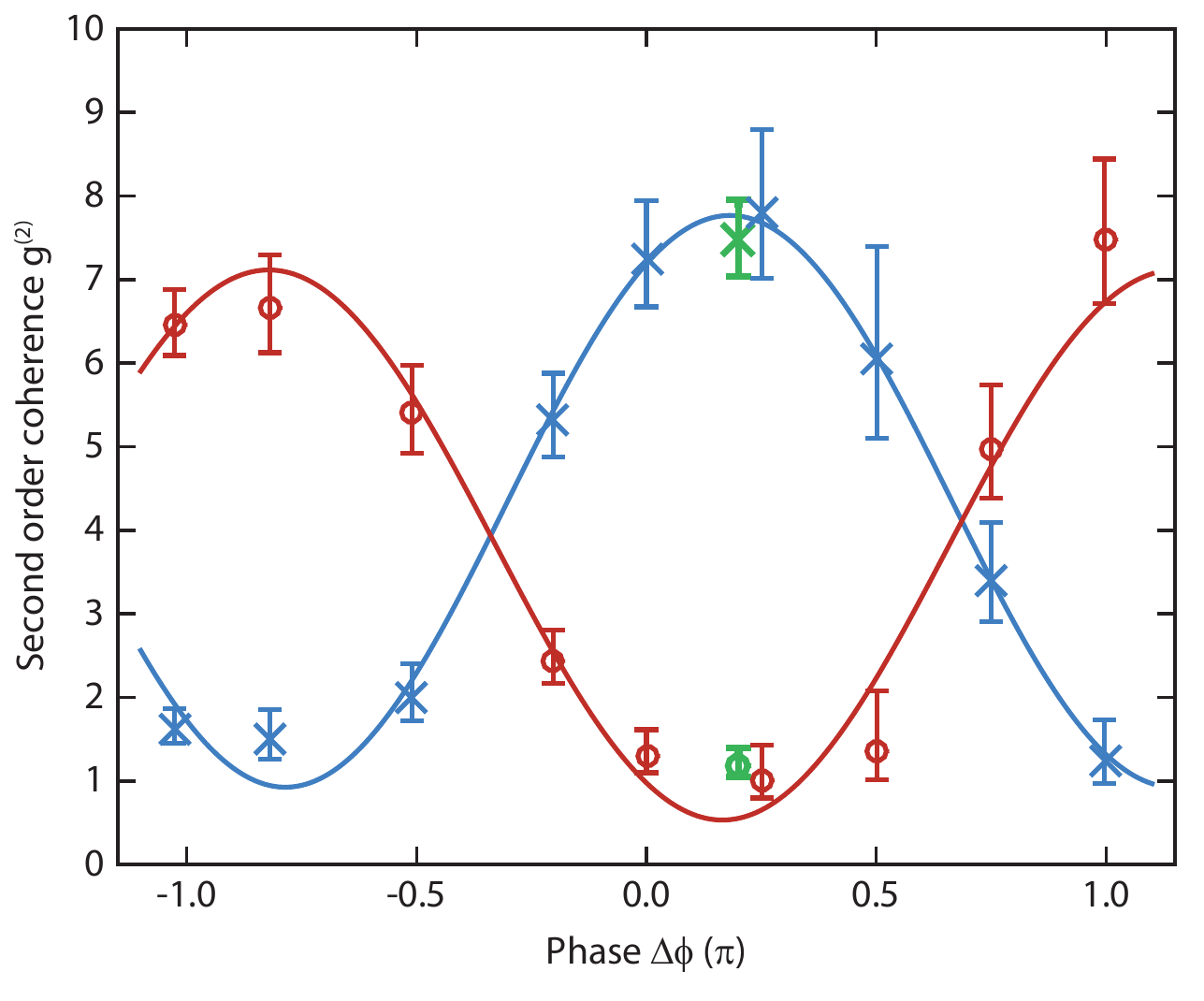}
		\caption{\textbf{Phase sweep of the entangled state.} We vary the phase difference between the pump and the read pulses, $\Delta\phi$, and measure the second-order coherence $g^{(2)}$ of the Raman-scattered photons for a fixed delay of	$\tau=123$~ns between the pulses. Blue crosses represent measurements of $g^{(2)}_{r_i,p_j,i\neq j}$ and red circles are the results for $g^{(2)}_{r_i,p_i}$, where $i,j\in\{1,2\}$. We fit simple sine functions (shown as solid lines) to each of the datasets as guides to the eye. The sinusoidal dependence on the phase clearly highlights the coherence of the entangled mechanical state. We observe a periodicity of 1.95$\pi$, in good agreement with the expected value of 2$\pi$ for single-particle interference (see equation~\eqref{eq:state})~\cite{Borkje2011}. The phase sweep allows us to identify the optimal phase $\Delta\phi=0.2\pi$ for maximum visibility, at which we acquire additional data (green cross and circle) to determine the entanglement witness with sufficient statistical significance. All error bars represent a 68\% confidence interval.}
		\label{fig:3}
	\end{center} 
\end{figure}

Although entanglement witnesses are designed to be efficient classifiers, they typically depend on the individual characteristics of the experimental setup. If, for example, the second beam splitter (see Fig.~\ref{fig:1}) were to malfunction and act as a perfect mirror--that is, if all photons from device A (B) were transmitted to detector 1 (2)--then $R_\textrm{m}(\theta ,j)$ could still be less than 1 for separable states. This is because the witness in equation~\eqref{eq:witness} estimates the visibility of the interference between $\ket{A}{}$ and $\ket{B}{}$ from a single measurement, without requiring a full phase scan of the interference fringe. To ensure the applicability of the witness, we therefore verify experimentally that our system fulfills its assumptions. We first check whether our setup is balanced by adjusting the energy of the pump pulses in each arm, as described above. This guarantees that the scattered photon fluxes impinging on the beam splitter from both arms are equal (see Supplementary Information). To make the detection symmetric, we use heralding detection events from both
superconducting nanowire single-photon detectors--that is, we obtain the actual bound on the entanglement witness $R_\textrm{m,sym}(\theta)$ from averaging measurements of $R_\textrm{m}(\theta, 1)$ and $R_\textrm{m}(\theta, 2)$ (see Supplementary Information). By choosing a phase $\theta$ such that the correlations between different detectors exceed the correlations at the same detector, $g^{(2)}_{r_i,p_j,i\neq j}>g^{(2)}_{r_i,p_i}$ with $i,j\in \{1,2\}$, we avoid our measurements' susceptibility to unequal splitting ratios applied by the beam splitter.

In Fig.~\ref{fig:3}, we show a series of measurements of the second-order coherence $g^{(2)}$, performed by sweeping $\Delta \phi$ with a readout delay of $\tau = 123$~ns, which verify the coherence between $|A\rangle$ and $|B\rangle$. Using these data, we chose an optimal phase setting $\theta=\theta_\textrm{opt}$ with $\Delta \phi= 0.2 \pi$ for the main experiment. We obtain $R_\textrm{m,sym}(\theta_\textrm{opt})=0.74^{+0.12}_{-0.06}$, which is well below the
separability bound of 1. By including measurements at the non-optimal adjacent phases $\Delta \phi=0$ and $0.25\pi$, the statistical uncertainty improves, and we obtain $R_\textrm{m,sym}([\theta_\textrm{opt}-0.2\pi, \theta_\textrm{opt} + 0.05\pi]) = 0.74^{+0.08}_{-0.05}$. Hence, we experimentally observe entanglement between the two remote
mechanical oscillators with a confidence level above 99.8\%.

The coherence properties of the generated state can be characterized through the decay of the visibility
\begin{equation}
V=\frac{\textrm{max}(g^{(2)}_{r_i,p_j})-\textrm{min}(g^{(2)}_{r_i,p_j})}{\textrm{max}(g^{(2)}_{r_i,p_j})+\textrm{min}(g^{(2)}_{r_i,p_j})}.
\end{equation}
We therefore sweep the delay time $\tau$ between the pump pulse and the read pulse. The mechanical frequency difference $\Delta \Omega_m$ allows us to sweep a full interference fringe by changing the delay $\tau$ by $22$~ns. Owing to the technically limited hold time of our cryostat, this sweep had to be performed at a higher bath temperature of about $80-90$~mK
(see Fig.~\ref{fig:1}), yielding a slightly lower, thermally limited visibility at short delays when compared to the data in Fig.~\ref{fig:3}. By varying the delay further, we observe interference between $\ket{A}{}$ and $\ket{B}{}$ ($V>0$) up to $\tau\approx3$~$\mu$s (see Fig.~\ref{fig:4}). The loss of coherence can be explained by absorption heating and mechanical decay (see Supplementary Information) and appears to be limited at long delays $\tau$ by the
lifetime $1/\Gamma_\textrm{A}\approx4$~$\mu$s of device A, which has the shorter lifetime of the two devices.

\begin{figure}[t]
	\begin{center}
		\includegraphics[width=0.9\columnwidth]{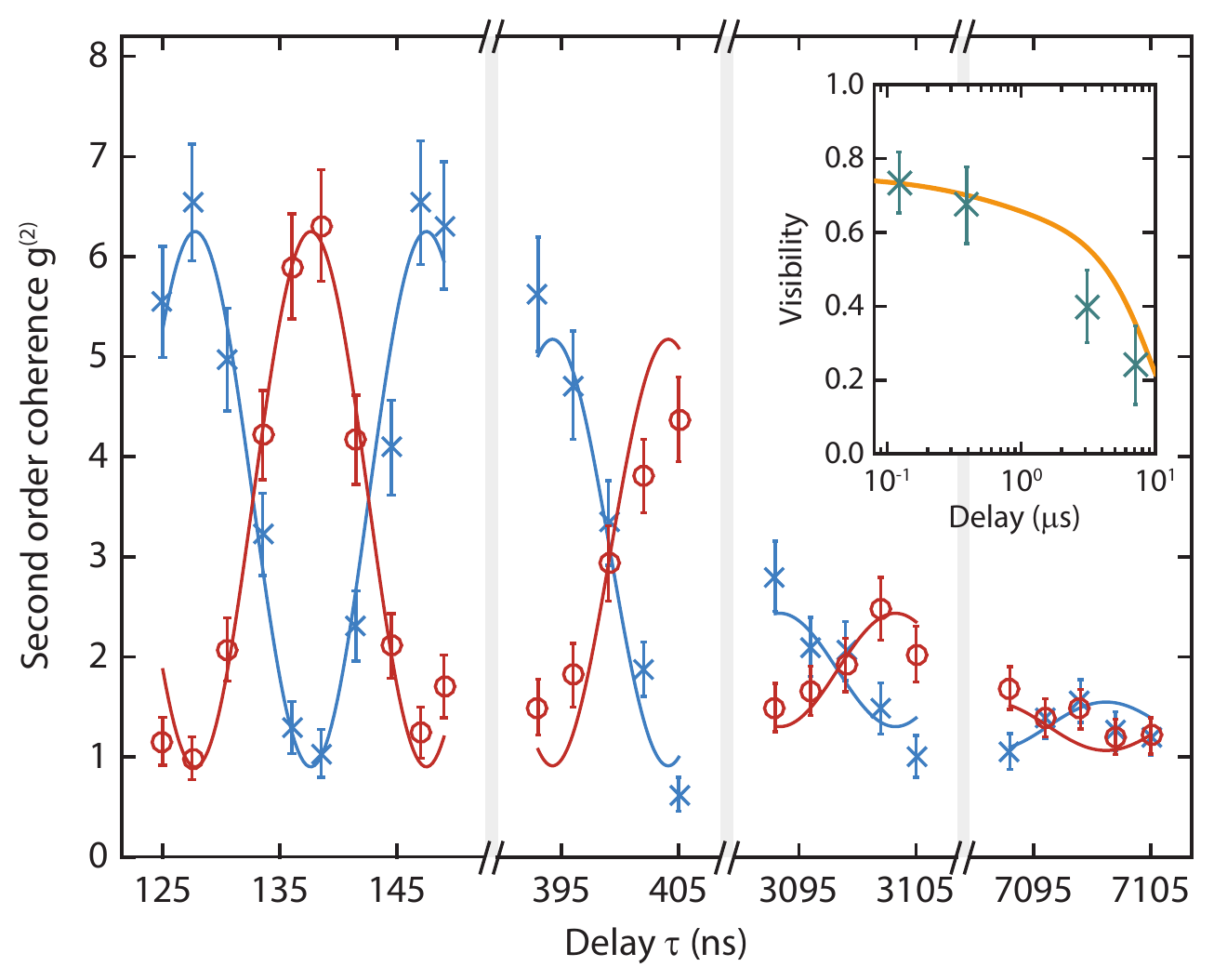}
		\caption{\textbf{Time sweep of the entangled state.} Shown is the interference of the entangled mechanical state at different delays $\tau$ between the pump and read pulses, with the phase of the interferometer, $\phi_0$, and the phase difference between the pump and read pulses, $\Delta\phi$, fixed. The blue crosses represent the measurements of $g^{(2)}_{r_i,p_j,i\neq j}$ and red circles are the results for $g^{(2)}_{r_i,p_i}$, where $i,j\in\{1,2\}$. The solid lines are sinusoidal fits averaged over the two out-of-phase components for each delay window and serve as a guide to the eye. The coherence of the entangled state is reduced over time, which can be seen by the decay of the interference visibility (inset). This decoherence is consistent with a delayed optical absorption heating and the mechanical decay time of about 4~$\mu$s of device A. The inset shows the visibility of the interference (green crosses) and the expected upper bound on the visibility due to heating and mechanical decay (orange line; see Supplementary Information). All error bars represent a 68\% confidence interval.}
		\label{fig:4}
	\end{center} 
\end{figure}

We have experimentally demonstrated entanglement between two engineered mechanical oscillators separated spatially by 20~cm and optically by 70~m. Imperfections in the fabrication process and the resulting small deviations of optical and mechanical frequencies for nominally identical devices are overcome through the statistical selection of devices and optical frequency shifting using a serrodyne approach. The mechanical systems do not interact directly at any point, but are interfaced remotely through optical photons in the telecommunication-wavelength band. The coherence time of the entangled state is several microseconds and appears to be limited by the mechanical lifetime of the devices and by absorption heating. Both of these limitations can be considerably mitigated. On the one hand, optical absorption can be substantially suppressed by using intrinsic, desiccated silicon~\cite{Asano2017}. Mechanical lifetimes, on the other hand, can be greatly increased by adding a phononic bandgap shield~\cite{Meenehan2015}. Although our devices are engineered to have short mechanical lifetimes~\cite{Hong2017,Patel2017}, earlier designs including such a phononic shield have reached~\cite{Meenehan2015} $1/\Gamma\approx0.5$~ms and could still be further improved. Combined with reduced optical absorption, which would allow efficient laser cooling, such lifetimes can potentially put our devices on par with other state-of-the-art quantum systems~\cite{Maring2017}.

Our experiment demonstrates a protocol for realistic, fibre telecommunication-compatible entanglement distribution using engineered mechanical quantum systems. With the current parameters of our system, a device separation of 75~km using commercially available telecommunication fibres would result in a drop of less than 5\% in the interference visibility (see discussion in Supplementary Information for more details). The system presented here is directly scalable to include more devices (see Supplementary Information) and could be integrated
into a real quantum network. Combining our results with those of optomechanical devices capable of transferring quantum information from the optical to the microwave domain, which is a highly active field of research~\cite{Bochmann2013,Rueda2016,Higginbotham2017}, could provide a backbone for a future quantum internet based on superconducting quantum computers.\\

\textbf{Acknowledgments}\ We would like to thank Vikas Anant, Klemens Hammerer, Joachim Hofer, Sebastian Hofer, Richard Norte, Kevin Phelan and Joshua Slater for valuable discussions and help. We also acknowledge assistance from the Kavli Nanolab Delft, in particular from Marc Zuiddam and Charles de Boer. This project was supported by the European Commission under the Marie Curie Horizon 2020 initial training programme OMT (grant 722923), Foundation for Fundamental Research on Matter (FOM) Projectruimte grants (15PR3210, 16PR1054), the Vienna Science and Technology Fund WWTF (ICT12-049), the European Research Council (ERC CoG QLev4G, ERC StG Strong-Q), the Austrian Science Fund (FWF) under projects F40 (SFB FOQUS) and P28172, and by the Netherlands Organisation for Scientific Research (NWO/OCW), as part of the Frontiers of Nanoscience program, as well as through a Vidi grant (680-47-541/994). R.R. is supported by the FWF under project W1210 (CoQuS) and is a recipient of a DOC fellowship of the Austrian Academy of Sciences at the University of Vienna.\\

\setcounter{figure}{0}
\renewcommand{\thefigure}{S\arabic{figure}}
\setcounter{equation}{0}
\renewcommand{\theequation}{S\arabic{equation}}

\clearpage

\section{Supplementary Information}

\subsection{Device fabrication and characterization}

\begin{figure}[t]
	\begin{center}
		\includegraphics[width=0.9\columnwidth]{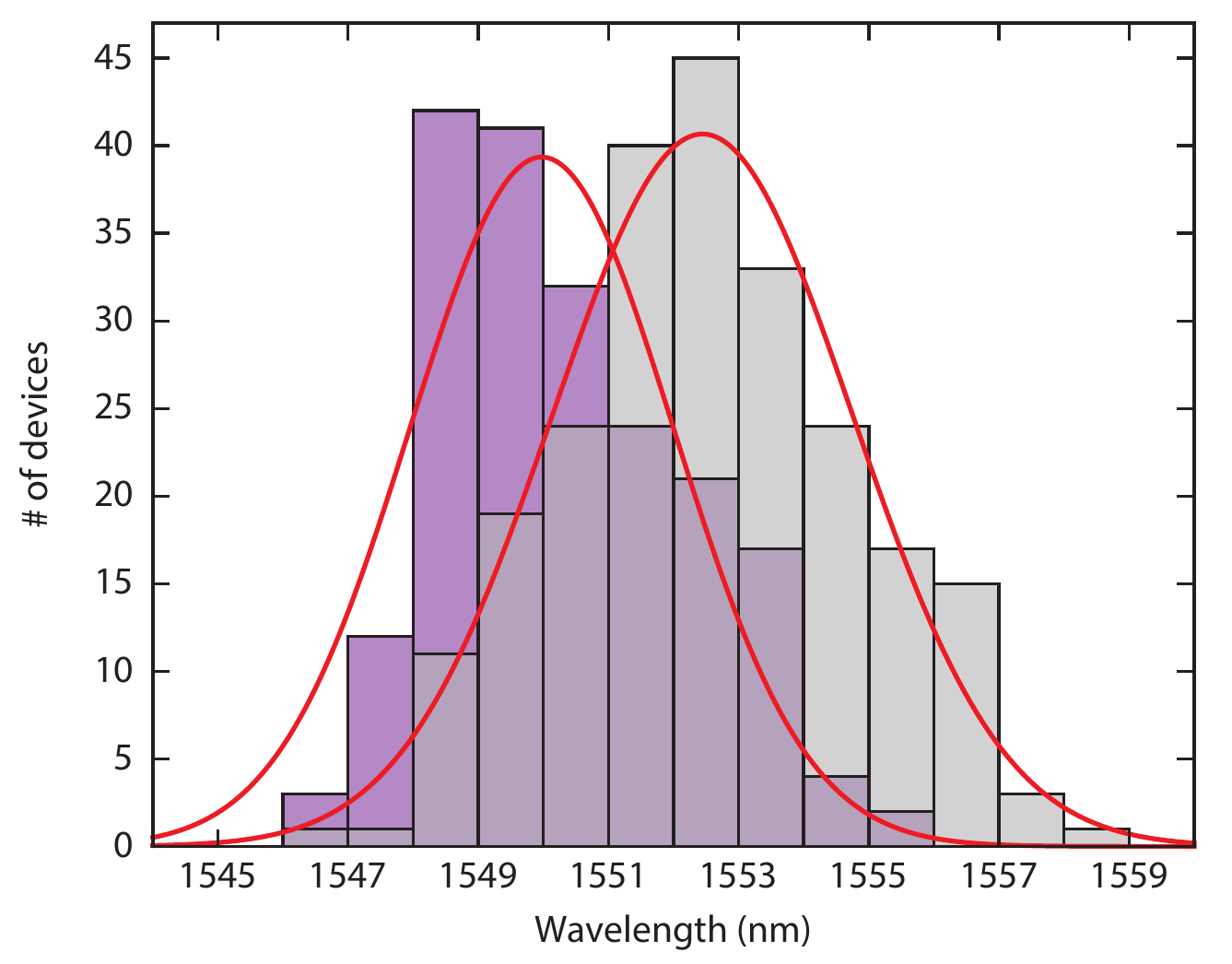}
		\caption{\textbf{Distribution of optical wavelengths.} We plot a histogram (bin size 1~nm) of the optical resonance wavelengths for a large set of devices on each of the chips containing devices A (gray) and B (magenta). The fits of a Gaussian distribution to the data sets (red solid lines) give a standard deviation of 2.3~nm and 2.0~nm, respectively. The large overlap of the optical resonance frequencies highlights the feasibility of extending the entanglement to even more optomechanical devices in the future.}
		\label{fig:S1}
	\end{center} 
\end{figure}

The devices in the main part are fabricated as described in reference~\cite{Hong2017}. The most crucial steps for generating two identical chips are the electron beam lithography and the inductively coupled plasma reactive ion etching. We beamwrite and etch on a single proto-chip containing two sets of devices. This chip is then diced into two halves, each with several hundred nominally identical resonators. The structures are subsequently released in 40\% hydrofluoric acid and cleaned with the RCA method, followed by a dip in 2\% hydrofluoric acid. When characterizing the two chips, we find the center wavelengths to be $1552.4$~nm on chip A and $1550.0$~nm on chip B (see Figure~\ref{fig:S1}). The standard deviation on the spread of the optical resonances is around $2$~nm on both chips. For the experiments in the main text, we search for resonances that overlap to within $10$\% of their linewidth, which is equal to around $100$~MHz. We find a total of $5$ pairs fulfilling this requirement within $234$ devices tested per chip.

In order to verify that finding identical devices is not just lucky coincidence and that this can even be done with a smaller sample size per chip, one can estimate the number of devices needed for a birthday paradox type approach. Therefore, we assume a pair of chips with $234$ devices each that are centered at the same target wavelength. Taking similar parameters as found in our actual chips, we use a spread in resonance wavelength of $2$~nm and we define resonances to be identical if they match to within $100$~MHz. While the probability of obtaining a single device exactly at the center wavelength is only 0.03\%, the probability of finding two matching devices at any wavelength within this distribution is $99.9996\%$. This is reduced if an offset in the mean wavelength of the two chips is introduced. For an offset of $2.5$~nm, the probability is $99.98$\%, and for $5$~nm, it is still $92.7$\%. By extending this approach to, for example, four chips, with the same parameters as above, no offset in the center wavelength and 500 devices per chip, we calculate the probability of finding four identical resonances to be 51.6\%. Such a quartet would directly enable experiments on entanglement swapping and tests of a Bell inequality, as proposed by DLCZ. Further improvements could include post-fabrication wavelength tuning, as has recently been demonstrated for similar devices~\cite{Fang2017}. This could significantly improve the prospect of scalability of our approach, as it would allow to fabricate identical devices more deterministically.

In addition, the mechanical resonances are also susceptible to fabrication errors and vary by up to 50~MHz for our devices. To overcome this mismatch, we use serrodyne frequency shifting (see sections below).

\subsection{Experimental setup}
\begin{figure*}[t]
	\begin{center}
		\includegraphics[width=0.9\textwidth]{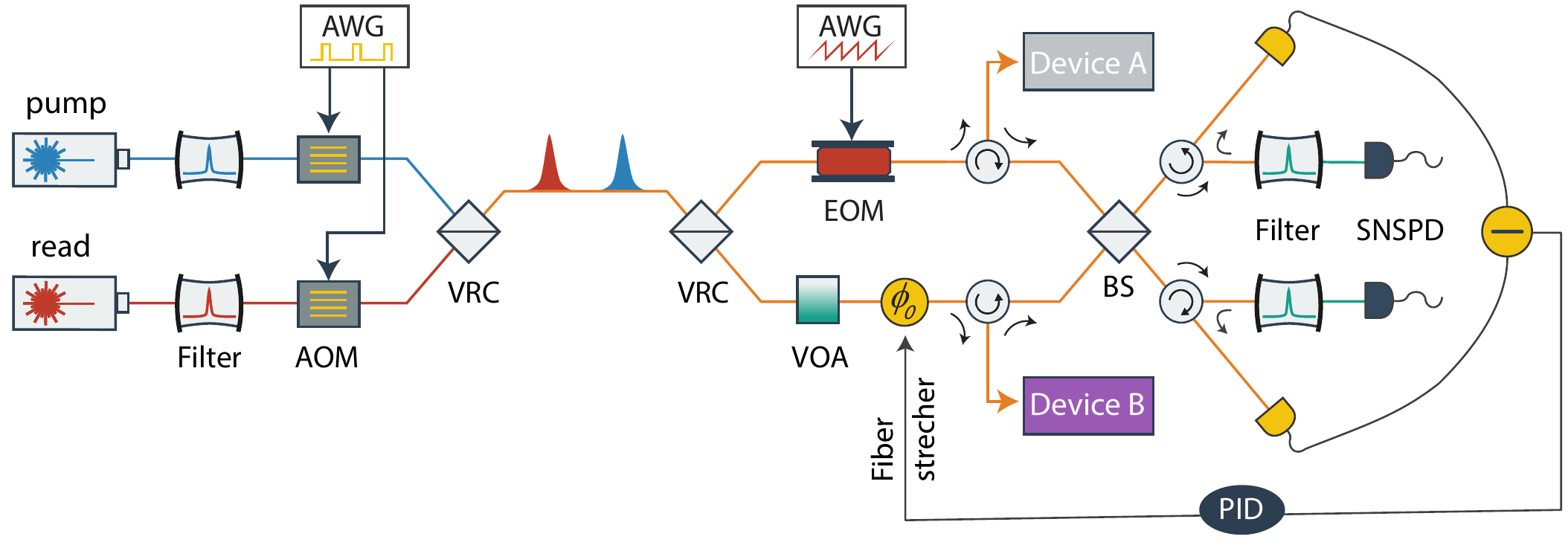}
		\caption{\textbf{Experimental setup.} A detailed schematics of our setup is shown here and described in the text. AOM are the acousto-optic modulators, AWG the arbitrary waveform generators, VRC the variable ratio couplers, EOM the electro-optic modulator, VOA the variable optical attenuator, BS the 50/50 beamsplitter and SNSPD the superconducting nanowire single-photon detectors.}
		\label{fig:S2}
	\end{center} 
\end{figure*}

A detailed drawing of the experimental setup is shown in Figure~\ref{fig:S2}. The light sources for our pump and read beams are two New Focus $6728$ CW lasers, tuned and stabilized on their respective sideband of the optical resonance. The beams are filtered by MicronOptics FFP-TF2 tunable optical filters in order to reduce the laser phase noise in the GHz regime. We then proceed to generate the actual pump and read pulses by driving acousto-optic modulators (Gooch\&Housego T-M110-0.2C2J-3-F2S) with an arbitrary function generator (Tektronic AFG3152C). These pulses are then combined on a variable ratio coupler (Newport F-CPL-1550-N-FA). The combined optical mode is subsequently split by another variable ratio coupler and fed into the Mach Zehnder interferometer. The coupling ratio is adjusted to primarily compensate for a small difference in total losses between two paths. The power in the interferometer arms can additionally be balanced by an electrically driven variable optical attenuator (Sercalo VP1). We reflect the pulses from the two devices via optical circulators and recombine on a $50/50$ coupler (measured deviation of $0.6$\%, see below). The strong pump pulses are filtered with two MicronOptics FFP-TF2 fiber filters per detection arm, tuned to transmit only the scattered (anti-) Stokes photons (bandwidth $50$ MHz). We detect the resonant photons with superconducting nanowire single photon detectors (Photonspot) and register their arrival times on a TimeHarp $260$ NANO correlation board.

\subsection{Serrodyne frequency shifting}

In our experiment, the mechanical frequency of device B ($\Omega_\textrm{m,B}$) is greater than of device A ($\Omega_\textrm{m,A}$) by $\Delta\Omega_\textrm{m}=2\pi\cdot45$~MHz. If we were to send pump pulses with exactly the same frequency $\omega_\textrm{pump}$ to both of the devices, they would produce scattered photons with frequencies $\omega_\textrm{o,A}=\omega_\textrm{pump}-\Omega_\textrm{m,A}$ and $\omega_\textrm{o,B}=\omega_\textrm{pump}-\Omega_\textrm{m,B}=\omega_\textrm{o,A}-\Delta\Omega_\textrm{m}$. This frequency mismatch of scattered photons from the two devices would make them distinguishable, therefore preventing the entangled state. A simple solution is to shift the frequency of the pump pulse going to the device A by $\Delta\Omega_\textrm{m}$, i.e.\ $\omega_\textrm{pump,A}=\omega_\textrm{pump}-\Delta\Omega_\textrm{m}$. We experimentally realize this by electrically driving the electro-optic phase modulator on the path to device A, with a sawtooth waveform. This so-called serrodyne modulation with frequency $\omega_\textrm{s}=\Delta\Omega_\textrm{m}$ and peak-to-peak phase amplitude of $2\pi$ results in an optical frequency shift of $\omega_\textrm{s}$~\cite{Cumming1957,Wong1982}. We use an arbitrary waveform generator (Agilent 81180A, bandwidth DC to 600~MHz) to generate the sawtooth voltage signal, amplify it with a broadband amplifier (Minicircuits TVA-R5-13A+, bandwidth 0.5 to 1000~MHz) and apply it to the optical phase modulator (Photline MPZ-LN-10-P-P-FA-FA-P, bandwidth DC to 12~GHz). We also apply an additional DC-bias to the serrodyne signal in order to generate a fixed phase offset $\Delta \phi$ in the interferometer arms. Due to the high analog bandwidth of the AWG compared to the frequency shift of 45~MHz, higher order sidebands were negligible and were not observed in the experiment.

\subsection{Phase stabilization of the interferometer}

For stabilizing the phase of the interferometer, we use an additional laser pulse $\sim$5~$\mu$s after the read pulse. To produce these auxiliary pulses, we also use the red-detuned laser that generates the read pulses and send them along the same beam paths. After being reflected from the optical filters in the detection line, the pulses are re-routed by optical circulators and picked up by a balanced detector (see Figure~\ref{fig:S2}). These signals are then sent to a PID controller, which regulates a fiber stretcher to stabilize the relative path length and therefore locking the phase of the interferometer on a slow timescale (i.e.\ with the experiment repetition period of 50~$\mu$s).

In principle, the read or pump pulses that are reflected off the filter cavities could also be used for the phase locking. However, the serrodyne modulation during pump and read results in a beat signal of the pulses behind the beam splitter. This beating requires more sophisticated signal processing, which we avoid by using the auxiliary pulses, during which the serrodyne modulation is off. We note that the auxiliary pulses also induce some absorption heating of the devices. However, the 50~$\mu$s repetition period is sufficiently long compared to the decay times of the devices for the extra heating not to influence our experimental result.

\subsection{Entanglement witness and Systematic Errors}

The entanglement witness~\cite{Horodecki2009}
\begin{equation}
R(\tau, j)=\frac{\left\langle  \hat m^\dagger_\textrm{A}(\tau) \hat m_\textrm{A}(\tau) \hat m^\dagger_\textrm{B}(\tau) \hat m_\textrm{B}(\tau) \right\rangle_j }{\left|\left\langle \hat m^\dagger_\textrm{A}(\tau) \hat m_\textrm{B}(\tau) \right\rangle_j\right|^2}
\label{eq:witness_source}
\end{equation}
derived in reference~\cite{Borkje2011} is based on the concurrence~\cite{Hill1997} of the bipartite mechanical system. Here, the conditional average $\langle \hat o \rangle_j=\langle \hat o \:\hat p^\dagger_j \hat p_j\rangle / \langle \hat p^\dagger_j \hat p_j\rangle$ for an operator $\hat o$ is its expectation value of the state heralded by a Stokes photon detected by detector $j$. The $\hat m_i$ ($\hat m^\dagger_i$) are the mechanical annihilation (creation) operators of device $i=\textrm{A},\textrm{B}$. While $R$ is experimentally not directly accessible, the upper bound to this witness $R_\textrm{m}$, see Eq.~\eqref{eq:witness}, is a measurable quantity in our interferometry setup. The derivation of the inequality $R_\textrm{m}\geq R$, as described in reference~\cite{Borkje2011} and its supplementary material, is based on several of assumptions:\ The unheralded state must be Gaussian at all times, and the interference on the combining beamsplitter must be symmetric. In this section, we would like to discuss the validity of each of these assumptions in more detail.

To obtain $R_\textrm{m}$, threefold coincidence measurements are re-expressed as twofold coincidences, which can be done for Gaussian states. Note that as the degrees of second order coherence in Eq.~\eqref{eq:witness} are measured between the pump and the read pulse, they are applied to this Gaussian state, not to the heralded, non-Gaussian entangled state $\ket{\Psi}{}\sim\ket{A}{}+e^{i\theta}\ket{B}{}$. We ensure that the mechanical states at the beginning of our protocol are Gaussian by allowing sufficiently long thermalization times (7x the mechanical decay time) prior to any optical manipulations. Consequently, the initial state is thermal, which for a bosonic system \cite{Riedinger2016,Hong2017} implies Gaussian quadrature statistics~\cite{Wieczorek2015}. Next, all optomechanical interactions involved in our protocol are linear~\cite{Hofer2011}, therefore conserving the Gaussianity of the state~\cite{Wieczorek2015}. Specifically, the Stokes process is described by the linear interaction Hamiltonian $\hat H_{S}\propto g_0 \hat m_i^\dagger \hat o_i^\dagger + \textrm{\textit{h.c.}}$ and the anti-Stokes process by $\hat H_{AS}\propto g_0 \hat m_i \hat o_i^\dagger + \textrm{\textit{h.c.}}$ for device $i=\textrm{A},\textrm{B}$ with the annihilation (creation) operator of optical resonance $\hat o_{i}$ ($\hat o_{i}^\dagger$).

Unintentional interactions, like absorption heating, happen probabilistically and in a remote frequency regime, such that it effectively acts as a Gaussian thermal bath~\cite{Meenehan2015,Hong2017}. Though not strictly contributing to the mechanical state, we also consider false positive detection events:\ drive photons leaking through the filter stages can be described by a coherent state (with Gaussian intensity fluctuations). Detection of stray photons and electrically caused false positive events are rare ($\sim$0.3\% of the total count rate in the detection window) and uncorrelated (autocorrelation $g^{(2)}(0)=1.05\pm 0.09$), such that it is reasonable to model them as a Gaussian process as well. More specifically, we observe an average added noise of $\sim$0.14 phonons during the measurement, from which the individual contributions of leaked drive photons, background detection events, and optical absorption heating can be estimated with additional measurements (for details see the section on "Second Order Coherence and Entanglement"). With the latter being a thermal process and therefore yielding Gaussian quadrature statistics, and an estimation of the initial thermal occupation from the nominal cryostat temperature, this leaves $\sim 4^{+2.8}_{-2.7}\cdot 10^{-2}$ phonons in device A and $\sim 0^{+1.4}_{-0}\cdot 10^{-2}$ in device B of unidentified origin. These are likely thermal phonons stemming from the non-ideal thermalization of the chips with their environment~\cite{Riedinger2016, Hong2017}.

For the modes leaving the interferometer \op{r}{j}, \op{p}{j}, reference~\cite{Borkje2011} assumes an ideal 50/50 beamsplitter with equal powers on each input. Small experimental deviations from this idealized scenario result in quadratic corrections to the witness. For example, when the ratio of read photons at the beamsplitter originating from devices A and B is $1+\delta$, the measurable upper bound changes to $R_\textrm{m} (1+\delta^2/2)\geq R$ for small $\delta$. In our experimental setup, we choose a fused fiber beamsplitter with a measured deviation of 0.6\%, leading to a relative correction on the order of $10^{-5}$. Experimentally, we cannot achieve power balancing at the input of the beamsplitter for photons scattered from the pump and the read pulses at the same time because the devices have slightly different thermal occupation. We choose to match the detection rates of heralding photons, i.e.\ photons scattered by the pump pulse. This preserves the unknown origin of the heralding photons. Differences in the optomechanical coupling strength and optical losses on the path from the device to the combining beamsplitter are compensated by adjusting the drive power in each path. After the balancing procedure, the relative difference in count rates of heralding photons is below 2\%, limited by the measurement precision during the balancing run and laser power fluctuations during the measurements. This leads to a relative correction of the witness below $10^{-3}$. The slightly different heating dynamics between devices A and B result in a measured flux ratio deviation of the scattered readout photons of $\delta \sim 5-10\%$. It can easily be seen from Equation~\eqref{eq:witness} that an increased heating will increase the witness $R_\textrm{m}$, and therefore the heating induced imbalance does not limit the validity of the witness. Yet, neglecting the thermal origin of the imbalancing, employing the correction $R_\textrm{m} (1+\delta^2/2)\geq R$, we obtain a relative systematic correction of $0.5\%$ to our result in the worst case scenario. Adding all systematic errors, we obtain a conservative upper bound of all relative systematic corrections of $\sim$0.5\%. Consequently, we use a reduced classicality bound of $R_m \geq 0.995$ instead of 1, reducing the confidence level slightly from $99.84\%$ to $99.82\%$.

\subsection{Statistical Analysis}

Results in the text and figures are given as maximum likelihood values and, where applicable, with a confidence interval of $\pm34\%$ around this value. For the second order coherences $g^{(2)}_{ri,pj},\;(i, j = 1, 2)$, we apply binomial statistics based on the number of counted two-fold coincidences, which dominates the statistical uncertainty~\cite{Riedinger2016}. The entanglement witness $R_\textrm{m}\left(\theta, j\right)$ in Eq.~\eqref{eq:witness} is a non-trivial function of multiple such $g^{(2)}_{ri,pj}$, expressed here as $R_\textrm{m}\left(\theta, j\right)\equiv\mathcal{R}_\textrm{m}\left(g^{(2)}_{r1,pj}\left(\theta\right), g^{(2)}_{r2,pj}\left(\theta\right)\right)$. To estimate its confidence intervals, we discretize the probability density function of the second order coherences $P(g^{(2)}_{ri,pj} \in [a; a + \delta a])$ at equidistant $a = n\delta a, n \in \mathbb{N}$. The probabilities for finding $R_\textrm{m}$ in an interval
$[f, f + \delta f]$ is then given by $P(R_\textrm{m} \in [f, f + \delta f]) = \sum_{(a,b)\in \mathcal{M}} P(g^{(2)}_{r1,pj} \in [a, a + \delta a])P(g^{(2)}_{r2,pj} \in [b, b + \delta b])$ on the set $\mathcal{M}$ for which $\mathcal{R}_\textrm{m}(a, b) \in [f, f + \delta f] \forall (a, b) \in \mathcal{M}$. For the optimal read phase $\theta_\textrm{opt}$ we obtain as maximum likelihood values for the witness bounds $R_\textrm{m}(\theta_\textrm{opt},1)=0.612^{+0.152}_{-0.057}$ and $R_\textrm{m}(\theta_\textrm{opt},2)=0.846^{+0.210}_{-0.090}$. We obtain the symmetrized witness by treating the experimentally observed $R_\textrm{m}(\theta_\textrm{opt},1)$ and $R_\textrm{m}(\theta_\textrm{opt},2)$ as two independent measurements of the expectation value $R_\textrm{m, sym} (\theta)\equiv \left\langle R_\textrm{m}(\theta,1) \right\rangle\stackrel{!}{=}\left\langle R_\textrm{m}(\theta,2) \right\rangle$ of the optomechanical state in a symmetric setup with no detector noise. For a realistic setup with detectors exhibiting different noise properties, without loss of generality, let detector $j$ have more false positive detection events than detector $i$. We then find the symmetrized witness $\left\langle R_\textrm{m}(\theta_\textrm{opt},i)\right\rangle \leq R_\textrm{m, sym} \leq \left\langle R_\textrm{m}(\theta_\textrm{opt},j)\right\rangle$ to be an upper bound to the entanglement witness of the state heralded by the better detector $i$. Consequently $1>R_\textrm{m, sym}\geq R_\textrm{m}(\theta_\textrm{opt},i)\geq R$ implies entanglement between the two remote mechanical oscillators. The observed values yield a confidence level for $R_\textrm{m, sym}<1$ of $98.4\%$. When correcting for the conservative upper bound of systematic errors, see above, the confidence level for having observed entanglement remains at $98.3\%$.

The complete counting statistics of the witness measurement at $\theta_\textrm{opt}$ are accumulated over $N = 1.114\cdot 10^9$ trials (i.e.\ sets of pulses). We obtain $C(p_1)=111134$ and $C(p_2)=184114$ counts for photons scattered by the pump pulse on detectors $i=1,2$ and $C(r_1)=108723$ and $C(r_2)=167427$ counts from the read pulse. This yields the coincidence counts $C_{ri,pj},\;(i, j = 1, 2)$ for counts on detector $i$, heralded by detector $j$,  $C_{r1,p1}=9$, $C_{r2,p1}=130$, $C_{r1,p2}=129$, $C_{r2,p2}=37$.

For non-optimal phases $\theta\neq\theta_\textrm{opt}$, $R_\textrm{m}(\theta,i)\geq R_\textrm{m}(\theta_\textrm{opt},i)$. Consequently, by adding the photon counts of measurements from an interval $[\theta_1, \theta_2]$, the resulting $R_\textrm{m}([\theta_1, \theta_2],i)\geq R_\textrm{m}(\theta \in [\theta_1, \theta_2],i)$ serves as an upper bound to any phase $\theta$ within that interval. Including the measurements at the non-optimal phase $\Delta \phi=0$, we obtain $R_\textrm{m}([\theta_\textrm{opt}-0.2\pi, \theta_\textrm{opt}],1)=0.66^{+0.114}_{-0.055}$ and $R_\textrm{m}([\theta_\textrm{opt}-0.2\pi, \theta_\textrm{opt}],2)=0.806^{+0.129}_{-0.071}$, resulting in $R_\textrm{m, sym}([\theta_\textrm{opt}-0.2\pi, \theta_\textrm{opt}])=0.74^{+0.08}_{-0.05} \geq R_\textrm{m, sym}(\theta_\textrm{opt})$.	The confidence level for $R_\textrm{m, sym}<1$ is $99.84\%$, dropping to $99.82\%$ when correcting for the conservative upper bound for systematic errors. Note that for states heralded only with the more efficient detector 1, we already have a confidence level for entanglement between the two mechanical oscillators of $99.50\%$, including corrections for systematic errors.

The statistics of the witness within the extended phase region are obtained from $N=1.949\cdot 10^9$ experimental trials. We get $C(p_1)=196080$ and $C(p_2)=322608$ counts for photons scattered by the pump pulse on detectors $i=1,2$ and $C(r_1)=194023$ and $C(r_2)=300373$ counts from the read pulse. This yields the coincidence counts $C_{ri,pj},\;(i, j = 1, 2)$ for counts on detector $i$, heralded by detector $j$,  $C_{r1,p1}=16$, $C_{r2,p1}=223$, $C_{r1,p2}=242$, $C_{r2,p2}=67$.

\subsection{Second Order Coherence and Entanglement}

The second order coherence between the scattered photons from the pump pulse and signal phonons transfered by the read pulse after a delay $t$, $g^{(2)}_{i,rp}(t)=\langle \opd{r}{j} \opd{p}{j} \op{r}{j} \op{p}{j}\rangle/\langle \opd{r}{j} \op{r}{j} \rangle \langle  \opd{p}{j} \op{p}{j}\rangle$, of the individual devices $i=A,B$, allows to us quantify the total noise contribution limiting the interference visibility. The measurements are performed the same way as described in the main text, however with the optical path to the other device blocked (see Fig.~\ref{fig:2}). Though there is no fundamental difference between the detectors $j=1,2$, we only use detector $j=2$ for the single device measurements. Following~\cite{Kuzmich2003} and starting from a thermal state for the mechanical system $\rho_m$ and vacuum $|0\rangle \langle 0|_o$ in the optical sidebands we obtain in the low temperature limit

\begin{equation}
g^{(2)}_{i,rp}(t) \approx 1+\frac{e^{-\Gamma_i t}}{n_{i,\textrm{th}}(t)+p_{\textrm{pump},i}\cdot e^{-\Gamma_i t}+n_\textrm{leak}+n_\textrm{bg}},
\label{eq:SIxcor}
\end{equation}

where $n_{i,\textrm{th}}\ll1$ is the mean phonon occupation of the device, $p_{\textrm{pump},i}\ll n_\textrm{th}$ is the Stokes excitation probability, $n_\textrm{leak}\ll n_\textrm{th}$ is the average number of leaked pump photons per transfered phonon, and $n_\textrm{bg}\ll n_\textrm{th}$ is the average number of background counts per detected phonon. At a delay of $\tau=123~\textrm{ns}\ll 1/\Gamma_i$ we measure $g^{(2)}_{\textrm{A,rp}}=7.1^{+1.2}_{-0.9}$ and $g^{(2)}_{\textrm{B,rp}}=9.6^{+1.1}_{-0.9}$.
With calibrated rates of $n_\textrm{bg}=3\cdot 10^{-3}$, $p_\textrm{pump,A}=0.56\cdot 10^{-2}$, $p_\textrm{pump,B}=0.80\cdot 10^{-2}$ and leaks at detector 1 $n_\textrm{leak,1}= 4.2 \cdot 10^{-2}$ and detector 2 $n_\textrm{leak,2}= 3.2 \cdot 10^{-2}$ we can estimate the number of incoherent phonons to be $n_\textrm{th,A}= 11.9^{+2.8}_{-2.7}\cdot 10^{-2}$ for device A and $n_\textrm{th,B}=6.9^{+1.4}_{-1.3}\cdot 10^{-2}$ for device B. In a more detailed analysis, including the mechanical decay measurement (see below), we can estimate that absorption by the pump pulse contributes $n_\textrm{pump}\sim 3\cdot 10^{-2}$ phonons and absorption from the read pulse contributes $n_\textrm{read}\sim 5\cdot 10^{-2}$ phonons. This suggest, that the performance of device A is limited by imperfect thermalization with the cryostat at a temperature of 60~mK.

\begin{figure*}[t]
	\begin{center}
		\includegraphics[width=0.9\textwidth]{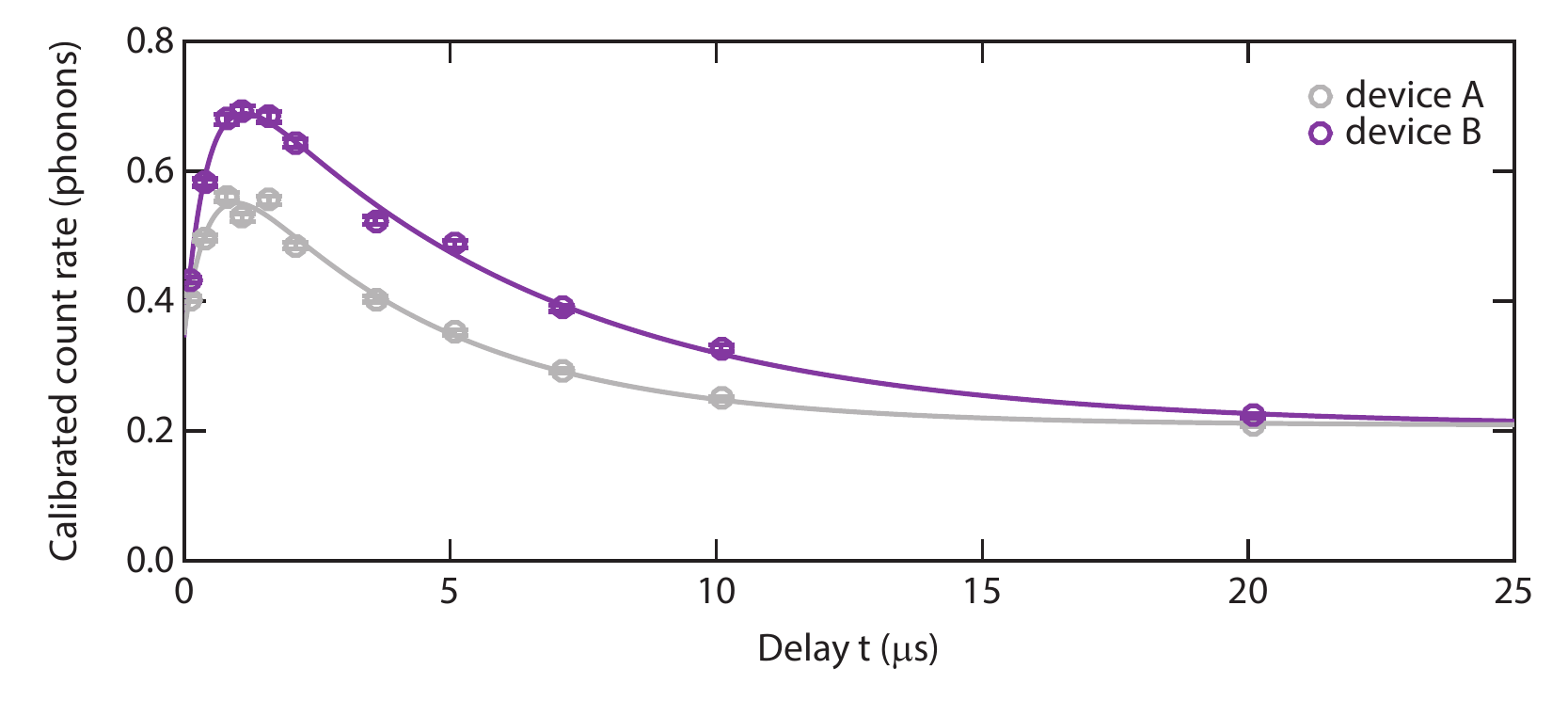}
		\caption{\textbf{Pump-probe experiment.} The experiment reveals the response of the devices mechanical modes to the initial optical pump. See text for details. Error bars are s.d.}
		\label{fig:S3}
	\end{center} 
\end{figure*}

The interference contrast of the entangled state, i.e.\ $C_\textrm{e}(t)=\textrm{max}(g^{(2)}_{ri,pj}(t,\theta))-\textrm{min}(g^{(2)}_{ri,pj}(t,\theta))$ is bound by the cross correlation of the individual devices $C_\textrm{e,max}(t)\approx \textrm{min}(g^{(2)}_{\textrm{A},rp}(t), g^{(2)}_{\textrm{B},rp}(t))-1$~\cite{LeePhD}. A pump-probe measurement of the thermal response of the devices~\cite{Riedinger2016} allows us to predict the cross correlations and thus an upper bound to the interference contrast. We excite the devices with a blue detuned pump pulse with $p_{\textrm{pump,A}}=2.8\%$ ($p_{\textrm{pump,B}}=4.0\%$) and vary the delay of a read (or probe) pulse with $5.6\%$ ($8.0\%$) state swap fidelity for device A (B). Note that for this experiment, we deliberately choose pulse energies higher than for the interference experiments in order to reduce the measurement time. As is done for sideband asymmetry~\cite{Riedinger2016,Hong2017}, the scattering rate of the probe pulse can be converted into the number of phonons at time $t$, when using the rate of the pump pulse (after scaling the signal according to the pump-to-probe power ratio) as a reference signal of single-phonon strength. This is only valid when $n_\textrm{th}\ll1$ at the time of pump pulse, which we find is the case later in the section. The results of these measurements are shown in Figure~\ref{fig:S3}. We observe an initial rise in phonon occupation after the pump pulse, followed by a decay to an equilibrium state. The delayed heating can be understood by the presence of long lived high frequency phonons, which weakly couple to the 5~GHz modes under investigation. We model these high frequency phonons by an effective thermal bath, exponentially decaying with rate $\gamma_i$. This results in the rate equation $\dot n_i(t)=-\Gamma_i n_i(t) + k_i e^{-\gamma_i t} + \Gamma_i n_{i,\textrm{init}}$ of the mean occupation number $n_i$ of device $i=\textrm{A,B}$. The additional bath couples with strength $k_i$, and the thermal environment of the chip has an equilibrium temperature of $n_{i,\textrm{init}}$. The detected signal of the probe pulse $d_i(t)=n_{i,\textrm{pump}} (t) + n_{i,\textrm{final}}$ contains the average phonon number $n_{i,\textrm{pump}} (t)=n_i(t)-n_{i,\textrm{init}}$ induced by the pump beam, as well as a constant offset $n_{i,\textrm{final}}=n_{i,\textrm{init}}+n_\textrm{i,probe}+n_\textrm{leak}+n_\textrm{bg}$ given by the thermal environment $n_{i,\textrm{init}}$, the false positive events $n_\textrm{leak}$ and $n_\textrm{bg}$ and the heating during the probe pulse itself $n_\textrm{i,probe}$. Consequently, we fit the pump-probe data with the general solution of the above differential equation $d_i(t)=a_i\cdot e^{-\Gamma_i t}-b_i\cdot e^{-\gamma_i t}+n_{i,\textrm{final}}$, with $a_i$, $b_i$ fitting parameters and including the offset $n_{i,\textrm{final}}$ of the probe pulse detection rate. We obtain $1/\Gamma_A=4.0~\mu$s, $1/\Gamma_B=5.8~\mu$s, $1/\gamma_A=0.5~\mu$s and $1/\gamma_B=0.5~\mu$s. Using the calibrated false positive detection rates and estimating the equilibrium occupation from the thermal environment $n_{i,\textrm{init}}\sim 1/(e^{\hbar\Omega/k_BT}-1)$, with the Boltzmann constant $k_B$ and cryostat temperature $T$, we can obtain the individual contributions to the absorption heating by the pump ($n_{i,\textrm{pump}} $) and the probe pulse ($n_{i,\textrm{probe}} $) for any given delay $t$. Adapting the number of phonons added by the optical drive pulses $n_{i,\textrm{pump}} $ and $n_{i,\textrm{probe}} $ for the lower energies, under the assumption of linear absorption and optomechanical processes, we obtain an estimate of the thermal occupation $n_{i,\textrm{th}}(t)=n_i(t)+n_{i,\textrm{probe}} -p_{\textrm{pump},i} e^{-\Gamma_i t}$ during the entanglement experiment. Using the calibrated rates and equation~\eqref{eq:SIxcor}, we can obtain an upper bound to the interference contrast $C_\textrm{e,max}$ (see above) and therefore also for the visibility $V_\textrm{max} = C_\textrm{e,max}/(C_\textrm{e,max}+2)$, which is shown in Figure~\ref{fig:4}.

\subsection{Rates and Extrapolation of Results}

In order to highlight the scalability of the mechanical entanglement, we recapitulate the detection rates in the entanglement witness measurement. The experiment is repeated every 50~$\mu$s, limited by the thermalization time of the mechanical modes. Using the counting statistics of the measurement at the optimal phase $\theta_\textrm{opt}$, see above, we have a probability of $p_\textrm{herald}\approx  2.7\cdot 10^{-4}$ and the unconditional probability to also detect the anti-Stokes photon from the readout pulse of $p_\textrm{read}\approx 2.8\cdot 10^{-7}$. These probabilities contain the optomechanically generated photons, leaked pump photons and background counts. In the current setup these rates are limited by losses in the filtering setup and low optomechanical scattering rates to retain the effects of absorption heating. When the two mechanical devices are placed in two separate refrigerators and placed in different locations, additional fiber would be inserted, causing additional losses. While the detection rate of leaked pump photons reduces in the same manner as the of the optomechanical photons, the rate of background counts stays the same. Consequently, the signal-to-noise ratio reduces, lowering the normalized cross-correlation between Stokes and anti-Stokes photons. In the present measurements, the inverse signal-to-background ratio is $n_{\textrm{bg},1}=2.9\cdot 10^{-3}$  ($n_{\textrm{bg},2}=3.2\cdot 10^{-3}$) for detectors 1 (2). The most reliable estimate of the second order coherence of device A, which has worse properties, can be extracted from the observed interference contrast, resulting in $g^{(2)}_{A,\textrm{rp}}\sim7.5\pm0.35$, as it has better statistics than the direct measurement. For device B, we use the direct measurement $g^{(2)}_{B,\textrm{rp}}=9.6^{+1.1}_{-0.9}$. To maintain an interference contrast of 95\% of the current level, the second order coherence of both devices is allowed to decrease to $\sim$7.1. With equation~\eqref{eq:SIxcor} we can estimate that an additional loss of 5.4~dB (10.6~dB) loss in the optical path from device A (B) to the combining beamsplitter decreases the signal-to-background ratio, such that $g^{(2)}_{A,rp}\approx g^{(2)}_{B,rp}\approx 7.1$. Using the nominal attenuation of commercial low-loss telecom fiber (Corning SMF-28 ULL $\sim$0.17~dB/km), we can estimate that an additional fiber length of $94^{+8}_{-12}$~km could be inserted between the devices. The projected entanglement witness in this configuration is $R_m\sim 0.76$ and would need roughly the same number of coincidence events to clear the classicality bound of 1 by 3 standard deviations. Including the reduced scattering rates from matching both paths, this would require $\sim$170~days of integration time, including a 15\% overhead time for stabilization of filters and interferometer as well as data management. Reducing the separation distance to 75~km, i.e.\ an additional fiber length of 32~km (43~km) between device A (B) and the beamsplitter, requires 38 days of continuous measurement time for a statistically significant demonstration of remote entanglement. While our cryostat currently only allows for a few weeks of measurements at a time due to technical limitations, much longer times should in principle be easily achievable.


\begin{thebibliography}{43}%
	\makeatletter
	\providecommand \@ifxundefined [1]{%
		\@ifx{#1\undefined}
	}%
	\providecommand \@ifnum [1]{%
		\ifnum #1\expandafter \@firstoftwo
		\else \expandafter \@secondoftwo
		\fi
	}%
	\providecommand \@ifx [1]{%
		\ifx #1\expandafter \@firstoftwo
		\else \expandafter \@secondoftwo
		\fi
	}%
	\providecommand \natexlab [1]{#1}%
	\providecommand \enquote  [1]{``#1''}%
	\providecommand \bibnamefont  [1]{#1}%
	\providecommand \bibfnamefont [1]{#1}%
	\providecommand \citenamefont [1]{#1}%
	\providecommand \href@noop [0]{\@secondoftwo}%
	\providecommand \href [0]{\begingroup \@sanitize@url \@href}%
	\providecommand \@href[1]{\@@startlink{#1}\@@href}%
	\providecommand \@@href[1]{\endgroup#1\@@endlink}%
	\providecommand \@sanitize@url [0]{\catcode `\\12\catcode `\$12\catcode
		`\&12\catcode `\#12\catcode `\^12\catcode `\_12\catcode `\%12\relax}%
	\providecommand \@@startlink[1]{}%
	\providecommand \@@endlink[0]{}%
	\providecommand \url  [0]{\begingroup\@sanitize@url \@url }%
	\providecommand \@url [1]{\endgroup\@href {#1}{\urlprefix }}%
	\providecommand \urlprefix  [0]{URL }%
	\providecommand \Eprint [0]{\href }%
	\providecommand \doibase [0]{http://dx.doi.org/}%
	\providecommand \selectlanguage [0]{\@gobble}%
	\providecommand \bibinfo  [0]{\@secondoftwo}%
	\providecommand \bibfield  [0]{\@secondoftwo}%
	\providecommand \translation [1]{[#1]}%
	\providecommand \BibitemOpen [0]{}%
	\providecommand \bibitemStop [0]{}%
	\providecommand \bibitemNoStop [0]{.\EOS\space}%
	\providecommand \EOS [0]{\spacefactor3000\relax}%
	\providecommand \BibitemShut  [1]{\csname bibitem#1\endcsname}%
	\let\auto@bib@innerbib\@empty
	\bibitem [{\citenamefont {Kimble}(2008)}]{Kimble2008}%
	\BibitemOpen
	\bibfield  {author} {\bibinfo {author} {\bibfnamefont {H.~J.}\ \bibnamefont
			{Kimble}},\ }\href {\doibase 10.1038/nature07127} {\bibfield  {journal}
		{\bibinfo  {journal} {Nature}\ }\textbf {\bibinfo {volume} {453}},\ \bibinfo
		{pages} {1023} (\bibinfo {year} {2008})}\BibitemShut {NoStop}%
	\bibitem [{\citenamefont {Jensen}\ \emph {et~al.}(2011)\citenamefont {Jensen},
		\citenamefont {Wasilewski}, \citenamefont {Krauter}, \citenamefont
		{Fernholz}, \citenamefont {Nielsen}, \citenamefont {Owari}, \citenamefont
		{Plenio}, \citenamefont {Serafini}, \citenamefont {Wolf},\ and\ \citenamefont
		{Polzik}}]{Jensen2011}%
	\BibitemOpen
	\bibfield  {author} {\bibinfo {author} {\bibfnamefont {K.}~\bibnamefont
			{Jensen}}, \bibinfo {author} {\bibfnamefont {W.}~\bibnamefont {Wasilewski}},
		\bibinfo {author} {\bibfnamefont {H.}~\bibnamefont {Krauter}}, \bibinfo
		{author} {\bibfnamefont {T.}~\bibnamefont {Fernholz}}, \bibinfo {author}
		{\bibfnamefont {B.~M.}\ \bibnamefont {Nielsen}}, \bibinfo {author}
		{\bibfnamefont {M.}~\bibnamefont {Owari}}, \bibinfo {author} {\bibfnamefont
			{M.~B.}\ \bibnamefont {Plenio}}, \bibinfo {author} {\bibfnamefont
			{A.}~\bibnamefont {Serafini}}, \bibinfo {author} {\bibfnamefont {M.~M.}\
			\bibnamefont {Wolf}}, \ and\ \bibinfo {author} {\bibfnamefont {E.~S.}\
			\bibnamefont {Polzik}},\ }\href {\doibase 10.1038/nphys1819} {\bibfield
		{journal} {\bibinfo  {journal} {Nature Phys.}\ }\textbf {\bibinfo {volume}
			{7}},\ \bibinfo {pages} {13} (\bibinfo {year} {2011})}\BibitemShut {NoStop}%
	\bibitem [{\citenamefont {Reim}\ \emph {et~al.}(2011)\citenamefont {Reim},
		\citenamefont {Michelberger}, \citenamefont {Lee}, \citenamefont {Nunn},
		\citenamefont {Langford},\ and\ \citenamefont {Walmsley}}]{Reim2011}%
	\BibitemOpen
	\bibfield  {author} {\bibinfo {author} {\bibfnamefont {K.~F.}\ \bibnamefont
			{Reim}}, \bibinfo {author} {\bibfnamefont {P.}~\bibnamefont {Michelberger}},
		\bibinfo {author} {\bibfnamefont {K.~C.}\ \bibnamefont {Lee}}, \bibinfo
		{author} {\bibfnamefont {J.}~\bibnamefont {Nunn}}, \bibinfo {author}
		{\bibfnamefont {N.~K.}\ \bibnamefont {Langford}}, \ and\ \bibinfo {author}
		{\bibfnamefont {I.~A.}\ \bibnamefont {Walmsley}},\ }\href {\doibase
		10.1103/PhysRevLett.107.053603} {\bibfield  {journal} {\bibinfo  {journal}
			{Phys.\ Rev.\ Lett.}\ }\textbf {\bibinfo {volume} {107}},\ \bibinfo {pages}
		{053603} (\bibinfo {year} {2011})}\BibitemShut {NoStop}%
	\bibitem [{\citenamefont {Chou}\ \emph {et~al.}(2005)\citenamefont {Chou},
		\citenamefont {de~Riedmatten}, \citenamefont {Felinto}, \citenamefont
		{Polyakov}, \citenamefont {van Enk},\ and\ \citenamefont
		{Kimble}}]{Chou2005}%
	\BibitemOpen
	\bibfield  {author} {\bibinfo {author} {\bibfnamefont {C.~W.}\ \bibnamefont
			{Chou}}, \bibinfo {author} {\bibfnamefont {H.}~\bibnamefont {de~Riedmatten}},
		\bibinfo {author} {\bibfnamefont {D.}~\bibnamefont {Felinto}}, \bibinfo
		{author} {\bibfnamefont {S.~V.}\ \bibnamefont {Polyakov}}, \bibinfo {author}
		{\bibfnamefont {S.~J.}\ \bibnamefont {van Enk}}, \ and\ \bibinfo {author}
		{\bibfnamefont {H.~J.}\ \bibnamefont {Kimble}},\ }\href {\doibase
		10.1038/nature04353} {\bibfield  {journal} {\bibinfo  {journal} {Nature}\
		}\textbf {\bibinfo {volume} {438}},\ \bibinfo {pages} {828} (\bibinfo {year}
		{2005})}\BibitemShut {NoStop}%
	\bibitem [{\citenamefont {Matsukevich}\ \emph {et~al.}(2006)\citenamefont
		{Matsukevich}, \citenamefont {Chanelie\`{e}re}, \citenamefont {Jenkins},
		\citenamefont {Lan}, \citenamefont {Kennedy},\ and\ \citenamefont
		{Kuzmich}}]{Matsukevich2006}%
	\BibitemOpen
	\bibfield  {author} {\bibinfo {author} {\bibfnamefont {D.~N.}\ \bibnamefont
			{Matsukevich}}, \bibinfo {author} {\bibfnamefont {T.}~\bibnamefont
			{Chanelie\`{e}re}}, \bibinfo {author} {\bibfnamefont {S.~D.}\ \bibnamefont
			{Jenkins}}, \bibinfo {author} {\bibfnamefont {S.-Y.}\ \bibnamefont {Lan}},
		\bibinfo {author} {\bibfnamefont {T.~A.~B.}\ \bibnamefont {Kennedy}}, \ and\
		\bibinfo {author} {\bibfnamefont {A.}~\bibnamefont {Kuzmich}},\ }\href
	{\doibase 10.1103/PhysRevLett.96.030405} {\bibfield  {journal} {\bibinfo
			{journal} {Phys.\ Rev.\ Lett.}\ }\textbf {\bibinfo {volume} {96}},\ \bibinfo
		{pages} {030405} (\bibinfo {year} {2006})}\BibitemShut {NoStop}%
	\bibitem [{\citenamefont {Ritter}\ \emph {et~al.}(2012)\citenamefont {Ritter},
		\citenamefont {N\"olleke}, \citenamefont {Hahn}, \citenamefont {Reiserer},
		\citenamefont {Neuzner}, \citenamefont {Uphoff}, \citenamefont {M\"ucke},
		\citenamefont {Figueroa}, \citenamefont {Bochmann},\ and\ \citenamefont
		{Rempe}}]{Ritter2012}%
	\BibitemOpen
	\bibfield  {author} {\bibinfo {author} {\bibfnamefont {S.}~\bibnamefont
			{Ritter}}, \bibinfo {author} {\bibfnamefont {C.}~\bibnamefont {N\"olleke}},
		\bibinfo {author} {\bibfnamefont {C.}~\bibnamefont {Hahn}}, \bibinfo {author}
		{\bibfnamefont {A.}~\bibnamefont {Reiserer}}, \bibinfo {author}
		{\bibfnamefont {A.}~\bibnamefont {Neuzner}}, \bibinfo {author} {\bibfnamefont
			{M.}~\bibnamefont {Uphoff}}, \bibinfo {author} {\bibfnamefont
			{M.}~\bibnamefont {M\"ucke}}, \bibinfo {author} {\bibfnamefont
			{E.}~\bibnamefont {Figueroa}}, \bibinfo {author} {\bibfnamefont
			{J.}~\bibnamefont {Bochmann}}, \ and\ \bibinfo {author} {\bibfnamefont
			{G.}~\bibnamefont {Rempe}},\ }\href {\doibase 10.1038/nature11023} {\bibfield
		{journal} {\bibinfo  {journal} {Nature}\ }\textbf {\bibinfo {volume}
			{484}},\ \bibinfo {pages} {195} (\bibinfo {year} {2012})}\BibitemShut
	{NoStop}%
	\bibitem [{\citenamefont {Moehring}\ \emph {et~al.}(2007)\citenamefont
		{Moehring}, \citenamefont {Maunz}, \citenamefont {Olmschenk}, \citenamefont
		{Younge}, \citenamefont {Matsukevich}, \citenamefont {Duan},\ and\
		\citenamefont {Monroe}}]{Moehring2007}%
	\BibitemOpen
	\bibfield  {author} {\bibinfo {author} {\bibfnamefont {D.~L.}\ \bibnamefont
			{Moehring}}, \bibinfo {author} {\bibfnamefont {P.}~\bibnamefont {Maunz}},
		\bibinfo {author} {\bibfnamefont {S.}~\bibnamefont {Olmschenk}}, \bibinfo
		{author} {\bibfnamefont {K.~C.}\ \bibnamefont {Younge}}, \bibinfo {author}
		{\bibfnamefont {D.~N.}\ \bibnamefont {Matsukevich}}, \bibinfo {author}
		{\bibfnamefont {L.-M.}\ \bibnamefont {Duan}}, \ and\ \bibinfo {author}
		{\bibfnamefont {C.}~\bibnamefont {Monroe}},\ }\href {\doibase
		10.1038/nature06118} {\bibfield  {journal} {\bibinfo  {journal} {Nature}\
		}\textbf {\bibinfo {volume} {449}},\ \bibinfo {pages} {68} (\bibinfo {year}
		{2007})}\BibitemShut {NoStop}%
	\bibitem [{\citenamefont {Jost}\ \emph {et~al.}(2009)\citenamefont {Jost},
		\citenamefont {Home}, \citenamefont {Amini}, \citenamefont {Hanneke},
		\citenamefont {Ozeri}, \citenamefont {Langer}, \citenamefont {Bollinger},
		\citenamefont {Leibfried},\ and\ \citenamefont {Wineland}}]{Jost2009}%
	\BibitemOpen
	\bibfield  {author} {\bibinfo {author} {\bibfnamefont {J.~D.}\ \bibnamefont
			{Jost}}, \bibinfo {author} {\bibfnamefont {J.~P.}\ \bibnamefont {Home}},
		\bibinfo {author} {\bibfnamefont {J.~M.}\ \bibnamefont {Amini}}, \bibinfo
		{author} {\bibfnamefont {D.}~\bibnamefont {Hanneke}}, \bibinfo {author}
		{\bibfnamefont {R.}~\bibnamefont {Ozeri}}, \bibinfo {author} {\bibfnamefont
			{C.}~\bibnamefont {Langer}}, \bibinfo {author} {\bibfnamefont {J.~J.}\
			\bibnamefont {Bollinger}}, \bibinfo {author} {\bibfnamefont {D.}~\bibnamefont
			{Leibfried}}, \ and\ \bibinfo {author} {\bibfnamefont {D.~J.}\ \bibnamefont
			{Wineland}},\ }\href {\doibase 10.1038/nature08006} {\bibfield  {journal}
		{\bibinfo  {journal} {Nature}\ }\textbf {\bibinfo {volume} {459}},\ \bibinfo
		{pages} {683} (\bibinfo {year} {2009})}\BibitemShut {NoStop}%
	\bibitem [{\citenamefont {Usmani}\ \emph {et~al.}(2012)\citenamefont {Usmani},
		\citenamefont {Clausen}, \citenamefont {Bussi\`{e}res}, \citenamefont
		{Sangouard}, \citenamefont {Afzelius},\ and\ \citenamefont
		{Gisin}}]{Usmani2012}%
	\BibitemOpen
	\bibfield  {author} {\bibinfo {author} {\bibfnamefont {I.}~\bibnamefont
			{Usmani}}, \bibinfo {author} {\bibfnamefont {C.}~\bibnamefont {Clausen}},
		\bibinfo {author} {\bibfnamefont {F.}~\bibnamefont {Bussi\`{e}res}}, \bibinfo
		{author} {\bibfnamefont {N.}~\bibnamefont {Sangouard}}, \bibinfo {author}
		{\bibfnamefont {M.}~\bibnamefont {Afzelius}}, \ and\ \bibinfo {author}
		{\bibfnamefont {N.}~\bibnamefont {Gisin}},\ }\href {\doibase
		10.1038/nphoton.2012.34} {\bibfield  {journal} {\bibinfo  {journal} {Nature
				Photon.}\ }\textbf {\bibinfo {volume} {6}},\ \bibinfo {pages} {234} (\bibinfo
		{year} {2012})}\BibitemShut {NoStop}%
	\bibitem [{\citenamefont {Saglamyurek}\ \emph {et~al.}(2015)\citenamefont
		{Saglamyurek}, \citenamefont {Jin}, \citenamefont {Verma}, \citenamefont
		{Shaw}, \citenamefont {Marsili}, \citenamefont {Nam}, \citenamefont {Oblak},\
		and\ \citenamefont {Tittel}}]{Saglamyurek2015}%
	\BibitemOpen
	\bibfield  {author} {\bibinfo {author} {\bibfnamefont {E.}~\bibnamefont
			{Saglamyurek}}, \bibinfo {author} {\bibfnamefont {J.}~\bibnamefont {Jin}},
		\bibinfo {author} {\bibfnamefont {V.~B.}\ \bibnamefont {Verma}}, \bibinfo
		{author} {\bibfnamefont {M.~D.}\ \bibnamefont {Shaw}}, \bibinfo {author}
		{\bibfnamefont {F.}~\bibnamefont {Marsili}}, \bibinfo {author} {\bibfnamefont
			{S.~W.}\ \bibnamefont {Nam}}, \bibinfo {author} {\bibfnamefont
			{D.}~\bibnamefont {Oblak}}, \ and\ \bibinfo {author} {\bibfnamefont
			{W.}~\bibnamefont {Tittel}},\ }\href {\doibase 10.1038/nphoton.2014.311}
	{\bibfield  {journal} {\bibinfo  {journal} {Nature Photon.}\ }\textbf
		{\bibinfo {volume} {9}},\ \bibinfo {pages} {83} (\bibinfo {year}
		{2015})}\BibitemShut {NoStop}%
	\bibitem [{\citenamefont {Hensen}\ \emph {et~al.}(2015)\citenamefont {Hensen},
		\citenamefont {Bernien}, \citenamefont {Dr{\'{e}}au}, \citenamefont
		{Reiserer}, \citenamefont {Kalb}, \citenamefont {Blok}, \citenamefont
		{Ruitenberg}, \citenamefont {Vermeulen}, \citenamefont {Schouten},
		\citenamefont {Abell{\'{a}}n}, \citenamefont {Amaya}, \citenamefont
		{Pruneri}, \citenamefont {Mitchell}, \citenamefont {Markham}, \citenamefont
		{Twitchen}, \citenamefont {Elkouss}, \citenamefont {Wehner}, \citenamefont
		{Taminiau},\ and\ \citenamefont {Hanson}}]{Hensen2015}%
	\BibitemOpen
	\bibfield  {author} {\bibinfo {author} {\bibfnamefont {B.}~\bibnamefont
			{Hensen}}, \bibinfo {author} {\bibfnamefont {H.}~\bibnamefont {Bernien}},
		\bibinfo {author} {\bibfnamefont {A.~E.}\ \bibnamefont {Dr{\'{e}}au}},
		\bibinfo {author} {\bibfnamefont {A.}~\bibnamefont {Reiserer}}, \bibinfo
		{author} {\bibfnamefont {N.}~\bibnamefont {Kalb}}, \bibinfo {author}
		{\bibfnamefont {M.~S.}\ \bibnamefont {Blok}}, \bibinfo {author}
		{\bibfnamefont {J.}~\bibnamefont {Ruitenberg}}, \bibinfo {author}
		{\bibfnamefont {R.~F.~L.}\ \bibnamefont {Vermeulen}}, \bibinfo {author}
		{\bibfnamefont {R.~N.}\ \bibnamefont {Schouten}}, \bibinfo {author}
		{\bibfnamefont {C.}~\bibnamefont {Abell{\'{a}}n}}, \bibinfo {author}
		{\bibfnamefont {W.}~\bibnamefont {Amaya}}, \bibinfo {author} {\bibfnamefont
			{V.}~\bibnamefont {Pruneri}}, \bibinfo {author} {\bibfnamefont {M.~W.}\
			\bibnamefont {Mitchell}}, \bibinfo {author} {\bibfnamefont {M.}~\bibnamefont
			{Markham}}, \bibinfo {author} {\bibfnamefont {D.~J.}\ \bibnamefont
			{Twitchen}}, \bibinfo {author} {\bibfnamefont {D.}~\bibnamefont {Elkouss}},
		\bibinfo {author} {\bibfnamefont {S.}~\bibnamefont {Wehner}}, \bibinfo
		{author} {\bibfnamefont {T.~H.}\ \bibnamefont {Taminiau}}, \ and\ \bibinfo
		{author} {\bibfnamefont {R.}~\bibnamefont {Hanson}},\ }\href {\doibase
		10.1038/nature15759} {\bibfield  {journal} {\bibinfo  {journal} {Nature}\
		}\textbf {\bibinfo {volume} {526}},\ \bibinfo {pages} {682} (\bibinfo {year}
		{2015})}\BibitemShut {NoStop}%
	\bibitem [{\citenamefont {Teufel}\ \emph {et~al.}(2011)\citenamefont {Teufel},
		\citenamefont {Donner}, \citenamefont {Li}, \citenamefont {Harlow},
		\citenamefont {Allman}, \citenamefont {Cicak}, \citenamefont {Sirois},
		\citenamefont {Whittaker}, \citenamefont {Lehnert},\ and\ \citenamefont
		{Simmonds}}]{Teufel2011b}%
	\BibitemOpen
	\bibfield  {author} {\bibinfo {author} {\bibfnamefont {J.~D.}\ \bibnamefont
			{Teufel}}, \bibinfo {author} {\bibfnamefont {T.}~\bibnamefont {Donner}},
		\bibinfo {author} {\bibfnamefont {D.}~\bibnamefont {Li}}, \bibinfo {author}
		{\bibfnamefont {J.~W.}\ \bibnamefont {Harlow}}, \bibinfo {author}
		{\bibfnamefont {M.~S.}\ \bibnamefont {Allman}}, \bibinfo {author}
		{\bibfnamefont {K.}~\bibnamefont {Cicak}}, \bibinfo {author} {\bibfnamefont
			{A.~J.}\ \bibnamefont {Sirois}}, \bibinfo {author} {\bibfnamefont {J.~D.}\
			\bibnamefont {Whittaker}}, \bibinfo {author} {\bibfnamefont {K.~W.}\
			\bibnamefont {Lehnert}}, \ and\ \bibinfo {author} {\bibfnamefont {R.~W.}\
			\bibnamefont {Simmonds}},\ }\href {\doibase 10.1038/nature10261} {\bibfield
		{journal} {\bibinfo  {journal} {Nature}\ }\textbf {\bibinfo {volume} {475}},\
		\bibinfo {pages} {359} (\bibinfo {year} {2011})}\BibitemShut {NoStop}%
	\bibitem [{\citenamefont {Chan}\ \emph {et~al.}(2011)\citenamefont {Chan},
		\citenamefont {Alegre}, \citenamefont {Safavi-Naeini}, \citenamefont {Hill},
		\citenamefont {Krause}, \citenamefont {Gr\"oblacher}, \citenamefont
		{Aspelmeyer},\ and\ \citenamefont {Painter}}]{Chan2011}%
	\BibitemOpen
	\bibfield  {author} {\bibinfo {author} {\bibfnamefont {J.}~\bibnamefont
			{Chan}}, \bibinfo {author} {\bibfnamefont {T.~P.~M.}\ \bibnamefont {Alegre}},
		\bibinfo {author} {\bibfnamefont {A.~H.}\ \bibnamefont {Safavi-Naeini}},
		\bibinfo {author} {\bibfnamefont {J.~T.}\ \bibnamefont {Hill}}, \bibinfo
		{author} {\bibfnamefont {A.}~\bibnamefont {Krause}}, \bibinfo {author}
		{\bibfnamefont {S.}~\bibnamefont {Gr\"oblacher}}, \bibinfo {author}
		{\bibfnamefont {M.}~\bibnamefont {Aspelmeyer}}, \ and\ \bibinfo {author}
		{\bibfnamefont {O.}~\bibnamefont {Painter}},\ }\href {\doibase
		10.1038/nature10461} {\bibfield  {journal} {\bibinfo  {journal} {Nature}\
		}\textbf {\bibinfo {volume} {478}},\ \bibinfo {pages} {89} (\bibinfo {year}
		{2011})}\BibitemShut {NoStop}%
	\bibitem [{\citenamefont {Palomaki}\ \emph {et~al.}(2013)\citenamefont
		{Palomaki}, \citenamefont {Teufel}, \citenamefont {Simmonds},\ and\
		\citenamefont {Lehnert}}]{Palomaki2013}%
	\BibitemOpen
	\bibfield  {author} {\bibinfo {author} {\bibfnamefont {T.}~\bibnamefont
			{Palomaki}}, \bibinfo {author} {\bibfnamefont {J.}~\bibnamefont {Teufel}},
		\bibinfo {author} {\bibfnamefont {R.}~\bibnamefont {Simmonds}}, \ and\
		\bibinfo {author} {\bibfnamefont {K.}~\bibnamefont {Lehnert}},\ }\href
	{\doibase 10.1126/science.1244563} {\bibfield  {journal} {\bibinfo  {journal}
			{Science}\ }\textbf {\bibinfo {volume} {342}},\ \bibinfo {pages} {710}
		(\bibinfo {year} {2013})}\BibitemShut {NoStop}%
	\bibitem [{\citenamefont {Riedinger}\ \emph {et~al.}(2016)\citenamefont
		{Riedinger}, \citenamefont {Hong}, \citenamefont {Norte}, \citenamefont
		{Slater}, \citenamefont {Shang}, \citenamefont {Krause}, \citenamefont
		{Anant}, \citenamefont {Aspelmeyer},\ and\ \citenamefont
		{Gr\"oblacher}}]{Riedinger2016}%
	\BibitemOpen
	\bibfield  {author} {\bibinfo {author} {\bibfnamefont {R.}~\bibnamefont
			{Riedinger}}, \bibinfo {author} {\bibfnamefont {S.}~\bibnamefont {Hong}},
		\bibinfo {author} {\bibfnamefont {R.~A.}\ \bibnamefont {Norte}}, \bibinfo
		{author} {\bibfnamefont {J.~A.}\ \bibnamefont {Slater}}, \bibinfo {author}
		{\bibfnamefont {J.}~\bibnamefont {Shang}}, \bibinfo {author} {\bibfnamefont
			{A.~G.}\ \bibnamefont {Krause}}, \bibinfo {author} {\bibfnamefont
			{V.}~\bibnamefont {Anant}}, \bibinfo {author} {\bibfnamefont
			{M.}~\bibnamefont {Aspelmeyer}}, \ and\ \bibinfo {author} {\bibfnamefont
			{S.}~\bibnamefont {Gr\"oblacher}},\ }\href {\doibase 10.1038/nature16536}
	{\bibfield  {journal} {\bibinfo  {journal} {Nature}\ }\textbf {\bibinfo
			{volume} {530}},\ \bibinfo {pages} {313} (\bibinfo {year}
		{2016})}\BibitemShut {NoStop}%
	\bibitem [{\citenamefont {Wollman}\ \emph {et~al.}(2015)\citenamefont
		{Wollman}, \citenamefont {Lei}, \citenamefont {Weinstein}, \citenamefont
		{Suh}, \citenamefont {Kronwald}, \citenamefont {Marquardt}, \citenamefont
		{Clerk},\ and\ \citenamefont {Schwab}}]{Wollman2015}%
	\BibitemOpen
	\bibfield  {author} {\bibinfo {author} {\bibfnamefont {E.~E.}\ \bibnamefont
			{Wollman}}, \bibinfo {author} {\bibfnamefont {C.~U.}\ \bibnamefont {Lei}},
		\bibinfo {author} {\bibfnamefont {A.~J.}\ \bibnamefont {Weinstein}}, \bibinfo
		{author} {\bibfnamefont {J.}~\bibnamefont {Suh}}, \bibinfo {author}
		{\bibfnamefont {A.}~\bibnamefont {Kronwald}}, \bibinfo {author}
		{\bibfnamefont {F.}~\bibnamefont {Marquardt}}, \bibinfo {author}
		{\bibfnamefont {A.~A.}\ \bibnamefont {Clerk}}, \ and\ \bibinfo {author}
		{\bibfnamefont {K.~C.}\ \bibnamefont {Schwab}},\ }\href {\doibase
		10.1126/science.aac5138} {\bibfield  {journal} {\bibinfo  {journal}
			{Science}\ }\textbf {\bibinfo {volume} {349}},\ \bibinfo {pages} {952}
		(\bibinfo {year} {2015})}\BibitemShut {NoStop}%
	\bibitem [{\citenamefont {Pirkkalainen}\ \emph {et~al.}(2015)\citenamefont
		{Pirkkalainen}, \citenamefont {Damsk\"agg}, \citenamefont {Brandt},
		\citenamefont {Massel},\ and\ \citenamefont
		{Sillanp\"a\"a}}]{Pirkkalainen2015}%
	\BibitemOpen
	\bibfield  {author} {\bibinfo {author} {\bibfnamefont {J.-M.}\ \bibnamefont
			{Pirkkalainen}}, \bibinfo {author} {\bibfnamefont {E.}~\bibnamefont
			{Damsk\"agg}}, \bibinfo {author} {\bibfnamefont {M.}~\bibnamefont {Brandt}},
		\bibinfo {author} {\bibfnamefont {F.}~\bibnamefont {Massel}}, \ and\ \bibinfo
		{author} {\bibfnamefont {M.~A.}\ \bibnamefont {Sillanp\"a\"a}},\ }\href
	{\doibase 10.1103/PhysRevLett.115.243601} {\bibfield  {journal} {\bibinfo
			{journal} {Phys.\ Rev.\ Lett.}\ }\textbf {\bibinfo {volume} {115}},\ \bibinfo
		{pages} {243601} (\bibinfo {year} {2015})}\BibitemShut {NoStop}%
	\bibitem [{\citenamefont {Lecocq}\ \emph {et~al.}(2015)\citenamefont {Lecocq},
		\citenamefont {Clark}, \citenamefont {Simmonds}, \citenamefont {Aumentado},\
		and\ \citenamefont {Teufel}}]{Lecocq2015}%
	\BibitemOpen
	\bibfield  {author} {\bibinfo {author} {\bibfnamefont {F.}~\bibnamefont
			{Lecocq}}, \bibinfo {author} {\bibfnamefont {J.~B.}\ \bibnamefont {Clark}},
		\bibinfo {author} {\bibfnamefont {R.~W.}\ \bibnamefont {Simmonds}}, \bibinfo
		{author} {\bibfnamefont {J.}~\bibnamefont {Aumentado}}, \ and\ \bibinfo
		{author} {\bibfnamefont {J.~D.}\ \bibnamefont {Teufel}},\ }\href {\doibase
		10.1103/PhysRevX.5.041037} {\bibfield  {journal} {\bibinfo  {journal} {Phys.\
				Rev.\ X}\ }\textbf {\bibinfo {volume} {5}},\ \bibinfo {pages} {041037}
		(\bibinfo {year} {2015})}\BibitemShut {NoStop}%
	\bibitem [{\citenamefont {O'Connell}\ \emph {et~al.}(2010)\citenamefont
		{O'Connell}, \citenamefont {Hofheinz}, \citenamefont {Ansmann}, \citenamefont
		{Bialczak}, \citenamefont {Lenander}, \citenamefont {Lucero}, \citenamefont
		{Neeley}, \citenamefont {Sank}, \citenamefont {Wang}, \citenamefont {Weides},
		\citenamefont {Wenner}, \citenamefont {Martinis},\ and\ \citenamefont
		{Cleland}}]{OConnell2010}%
	\BibitemOpen
	\bibfield  {author} {\bibinfo {author} {\bibfnamefont {A.~D.}\ \bibnamefont
			{O'Connell}}, \bibinfo {author} {\bibfnamefont {M.}~\bibnamefont {Hofheinz}},
		\bibinfo {author} {\bibfnamefont {M.}~\bibnamefont {Ansmann}}, \bibinfo
		{author} {\bibfnamefont {R.~C.}\ \bibnamefont {Bialczak}}, \bibinfo {author}
		{\bibfnamefont {M.}~\bibnamefont {Lenander}}, \bibinfo {author}
		{\bibfnamefont {E.}~\bibnamefont {Lucero}}, \bibinfo {author} {\bibfnamefont
			{M.}~\bibnamefont {Neeley}}, \bibinfo {author} {\bibfnamefont
			{D.}~\bibnamefont {Sank}}, \bibinfo {author} {\bibfnamefont {H.}~\bibnamefont
			{Wang}}, \bibinfo {author} {\bibfnamefont {M.}~\bibnamefont {Weides}},
		\bibinfo {author} {\bibfnamefont {J.}~\bibnamefont {Wenner}}, \bibinfo
		{author} {\bibfnamefont {J.~M.}\ \bibnamefont {Martinis}}, \ and\ \bibinfo
		{author} {\bibfnamefont {A.~N.}\ \bibnamefont {Cleland}},\ }\href {\doibase
		10.1038/nature08967} {\bibfield  {journal} {\bibinfo  {journal} {Nature}\
		}\textbf {\bibinfo {volume} {464}},\ \bibinfo {pages} {697} (\bibinfo {year}
		{2010})}\BibitemShut {NoStop}%
	\bibitem [{\citenamefont {Chu}\ \emph {et~al.}(2017)\citenamefont {Chu},
		\citenamefont {Kharel}, \citenamefont {Renninger}, \citenamefont {Burkhart},
		\citenamefont {Frunzio}, \citenamefont {Rakich},\ and\ \citenamefont
		{Schoelkopf}}]{Chu2017}%
	\BibitemOpen
	\bibfield  {author} {\bibinfo {author} {\bibfnamefont {Y.}~\bibnamefont
			{Chu}}, \bibinfo {author} {\bibfnamefont {P.}~\bibnamefont {Kharel}},
		\bibinfo {author} {\bibfnamefont {W.~H.}\ \bibnamefont {Renninger}}, \bibinfo
		{author} {\bibfnamefont {L.~D.}\ \bibnamefont {Burkhart}}, \bibinfo {author}
		{\bibfnamefont {L.}~\bibnamefont {Frunzio}}, \bibinfo {author} {\bibfnamefont
			{P.~T.}\ \bibnamefont {Rakich}}, \ and\ \bibinfo {author} {\bibfnamefont
			{R.~J.}\ \bibnamefont {Schoelkopf}},\ }\href {\doibase
		10.1126/science.aao1511} {\bibfield  {journal} {\bibinfo  {journal}
			{Science}\ }\textbf {\bibinfo {volume} {358}},\ \bibinfo {pages} {199}
		(\bibinfo {year} {2017})}\BibitemShut {NoStop}%
	\bibitem [{\citenamefont {Hong}\ \emph {et~al.}(2017)\citenamefont {Hong},
		\citenamefont {Riedinger}, \citenamefont {Marinkovi\'{c}}, \citenamefont
		{Wallucks}, \citenamefont {Hofer}, \citenamefont {Norte}, \citenamefont
		{Aspelmeyer},\ and\ \citenamefont {Gr\"oblacher}}]{Hong2017}%
	\BibitemOpen
	\bibfield  {author} {\bibinfo {author} {\bibfnamefont {S.}~\bibnamefont
			{Hong}}, \bibinfo {author} {\bibfnamefont {R.}~\bibnamefont {Riedinger}},
		\bibinfo {author} {\bibfnamefont {I.}~\bibnamefont {Marinkovi\'{c}}},
		\bibinfo {author} {\bibfnamefont {A.}~\bibnamefont {Wallucks}}, \bibinfo
		{author} {\bibfnamefont {S.~G.}\ \bibnamefont {Hofer}}, \bibinfo {author}
		{\bibfnamefont {R.~A.}\ \bibnamefont {Norte}}, \bibinfo {author}
		{\bibfnamefont {M.}~\bibnamefont {Aspelmeyer}}, \ and\ \bibinfo {author}
		{\bibfnamefont {S.}~\bibnamefont {Gr\"oblacher}},\ }\href {\doibase
		10.1126/science.aan7939} {\bibfield  {journal} {\bibinfo  {journal}
			{Science}\ }\textbf {\bibinfo {volume} {358}},\ \bibinfo {pages} {203}
		(\bibinfo {year} {2017})}\BibitemShut {NoStop}%
	\bibitem [{\citenamefont {Reed}\ \emph {et~al.}(2017)\citenamefont {Reed},
		\citenamefont {Mayer}, \citenamefont {Teufel}, \citenamefont {Burkhart},
		\citenamefont {Pfaff}, \citenamefont {Reagor}, \citenamefont {Sletten},
		\citenamefont {Ma}, \citenamefont {Schoelkopf}, \citenamefont {Knill},\ and\
		\citenamefont {Lehnert}}]{Reed2017}%
	\BibitemOpen
	\bibfield  {author} {\bibinfo {author} {\bibfnamefont {A.~P.}\ \bibnamefont
			{Reed}}, \bibinfo {author} {\bibfnamefont {K.~H.}\ \bibnamefont {Mayer}},
		\bibinfo {author} {\bibfnamefont {J.~D.}\ \bibnamefont {Teufel}}, \bibinfo
		{author} {\bibfnamefont {L.~D.}\ \bibnamefont {Burkhart}}, \bibinfo {author}
		{\bibfnamefont {W.}~\bibnamefont {Pfaff}}, \bibinfo {author} {\bibfnamefont
			{M.}~\bibnamefont {Reagor}}, \bibinfo {author} {\bibfnamefont
			{L.}~\bibnamefont {Sletten}}, \bibinfo {author} {\bibfnamefont
			{X.}~\bibnamefont {Ma}}, \bibinfo {author} {\bibfnamefont {R.~J.}\
			\bibnamefont {Schoelkopf}}, \bibinfo {author} {\bibfnamefont
			{E.}~\bibnamefont {Knill}}, \ and\ \bibinfo {author} {\bibfnamefont {K.~W.}\
			\bibnamefont {Lehnert}},\ }\href {\doibase 10.1038/nphys4251} {\bibfield
		{journal} {\bibinfo  {journal} {Nature Phys.}\ }\textbf {\bibinfo {volume}
			{13}},\ \bibinfo {pages} {1163} (\bibinfo {year} {2017})}\BibitemShut
	{NoStop}%
	\bibitem [{\citenamefont {Lee}\ \emph {et~al.}(2011)\citenamefont {Lee},
		\citenamefont {Sprague}, \citenamefont {Sussman}, \citenamefont {Nunn},
		\citenamefont {Langford}, \citenamefont {Jin}, \citenamefont {Champion},
		\citenamefont {Michelberger}, \citenamefont {Reim}, \citenamefont {England},
		\citenamefont {Jaksch},\ and\ \citenamefont {Walmsley}}]{Lee2011}%
	\BibitemOpen
	\bibfield  {author} {\bibinfo {author} {\bibfnamefont {K.~C.}\ \bibnamefont
			{Lee}}, \bibinfo {author} {\bibfnamefont {M.~R.}\ \bibnamefont {Sprague}},
		\bibinfo {author} {\bibfnamefont {B.~J.}\ \bibnamefont {Sussman}}, \bibinfo
		{author} {\bibfnamefont {J.}~\bibnamefont {Nunn}}, \bibinfo {author}
		{\bibfnamefont {N.~K.}\ \bibnamefont {Langford}}, \bibinfo {author}
		{\bibfnamefont {X.-M.}\ \bibnamefont {Jin}}, \bibinfo {author} {\bibfnamefont
			{T.}~\bibnamefont {Champion}}, \bibinfo {author} {\bibfnamefont
			{P.}~\bibnamefont {Michelberger}}, \bibinfo {author} {\bibfnamefont {K.~F.}\
			\bibnamefont {Reim}}, \bibinfo {author} {\bibfnamefont {D.}~\bibnamefont
			{England}}, \bibinfo {author} {\bibfnamefont {D.}~\bibnamefont {Jaksch}}, \
		and\ \bibinfo {author} {\bibfnamefont {I.}~\bibnamefont {Walmsley}},\ }\href
	{\doibase 10.1126/science.1211914} {\bibfield  {journal} {\bibinfo  {journal}
			{Science}\ }\textbf {\bibinfo {volume} {334}},\ \bibinfo {pages} {1253}
		(\bibinfo {year} {2011})}\BibitemShut {NoStop}%
	\bibitem [{\citenamefont {Meenehan}\ \emph {et~al.}(2015)\citenamefont
		{Meenehan}, \citenamefont {Cohen}, \citenamefont {MacCabe}, \citenamefont
		{Marsili}, \citenamefont {Shaw},\ and\ \citenamefont
		{Painter}}]{Meenehan2015}%
	\BibitemOpen
	\bibfield  {author} {\bibinfo {author} {\bibfnamefont {S.~M.}\ \bibnamefont
			{Meenehan}}, \bibinfo {author} {\bibfnamefont {J.~D.}\ \bibnamefont {Cohen}},
		\bibinfo {author} {\bibfnamefont {G.~S.}\ \bibnamefont {MacCabe}}, \bibinfo
		{author} {\bibfnamefont {F.}~\bibnamefont {Marsili}}, \bibinfo {author}
		{\bibfnamefont {M.~D.}\ \bibnamefont {Shaw}}, \ and\ \bibinfo {author}
		{\bibfnamefont {O.}~\bibnamefont {Painter}},\ }\href {\doibase
		10.1103/PhysRevX.5.041002} {\bibfield  {journal} {\bibinfo  {journal} {Phys.\
				Rev.\ X}\ }\textbf {\bibinfo {volume} {5}},\ \bibinfo {pages} {041002}
		(\bibinfo {year} {2015})}\BibitemShut {NoStop}%
	\bibitem [{\citenamefont {Razavi}\ \emph {et~al.}(2009)\citenamefont {Razavi},
		\citenamefont {Piani},\ and\ \citenamefont {L\"utkenhaus}}]{Razavi2009}%
	\BibitemOpen
	\bibfield  {author} {\bibinfo {author} {\bibfnamefont {M.}~\bibnamefont
			{Razavi}}, \bibinfo {author} {\bibfnamefont {M.}~\bibnamefont {Piani}}, \
		and\ \bibinfo {author} {\bibfnamefont {N.}~\bibnamefont {L\"utkenhaus}},\
	}\href {\doibase 10.1103/PhysRevA.80.032301} {\bibfield  {journal} {\bibinfo
			{journal} {Phys.\ Rev.\ A}\ }\textbf {\bibinfo {volume} {80}},\ \bibinfo
		{pages} {032301} (\bibinfo {year} {2009})}\BibitemShut {NoStop}%
	\bibitem [{\citenamefont {Bochmann}\ \emph {et~al.}(2013)\citenamefont
		{Bochmann}, \citenamefont {Vainsencher}, \citenamefont {Awschalom},\ and\
		\citenamefont {Cleland}}]{Bochmann2013}%
	\BibitemOpen
	\bibfield  {author} {\bibinfo {author} {\bibfnamefont {J.}~\bibnamefont
			{Bochmann}}, \bibinfo {author} {\bibfnamefont {A.}~\bibnamefont
			{Vainsencher}}, \bibinfo {author} {\bibfnamefont {D.~D.}\ \bibnamefont
			{Awschalom}}, \ and\ \bibinfo {author} {\bibfnamefont {A.~N.}\ \bibnamefont
			{Cleland}},\ }\href {\doibase 10.1038/nphys2748} {\bibfield  {journal}
		{\bibinfo  {journal} {Nature Phys.}\ }\textbf {\bibinfo {volume} {9}},\
		\bibinfo {pages} {712} (\bibinfo {year} {2013})}\BibitemShut {NoStop}%
	\bibitem [{\citenamefont {Duan}\ \emph {et~al.}(2001)\citenamefont {Duan},
		\citenamefont {Lukin}, \citenamefont {Cirac},\ and\ \citenamefont
		{Zoller}}]{Duan2001}%
	\BibitemOpen
	\bibfield  {author} {\bibinfo {author} {\bibfnamefont {L.~M.}\ \bibnamefont
			{Duan}}, \bibinfo {author} {\bibfnamefont {M.~D.}\ \bibnamefont {Lukin}},
		\bibinfo {author} {\bibfnamefont {J.~I.}\ \bibnamefont {Cirac}}, \ and\
		\bibinfo {author} {\bibfnamefont {P.}~\bibnamefont {Zoller}},\ }\href
	{\doibase 10.1038/35106500} {\bibfield  {journal} {\bibinfo  {journal}
			{Nature}\ }\textbf {\bibinfo {volume} {414}},\ \bibinfo {pages} {413}
		(\bibinfo {year} {2001})}\BibitemShut {NoStop}%
	\bibitem [{\citenamefont {Chan}(2012)}]{ChanPhD}%
	\BibitemOpen
	\bibfield  {author} {\bibinfo {author} {\bibfnamefont {J.}~\bibnamefont
			{Chan}},\ }\emph {\bibinfo {title} {Laser cooling of an optomechanical
			crystal resonator to its quantum ground state of motion}},\ \href
	{https://thesis.library.caltech.edu/7098/} {Ph.D. thesis},\ \bibinfo
	{school} {California Institute of Technology} (\bibinfo {year}
	{2012})\BibitemShut {NoStop}%
	\bibitem [{\citenamefont {B\o{}rkje}\ \emph {et~al.}(2011)\citenamefont
		{B\o{}rkje}, \citenamefont {Nunnenkamp},\ and\ \citenamefont
		{Girvin}}]{Borkje2011}%
	\BibitemOpen
	\bibfield  {author} {\bibinfo {author} {\bibfnamefont {K.}~\bibnamefont
			{B\o{}rkje}}, \bibinfo {author} {\bibfnamefont {A.}~\bibnamefont
			{Nunnenkamp}}, \ and\ \bibinfo {author} {\bibfnamefont {S.~M.}\ \bibnamefont
			{Girvin}},\ }\href {\doibase 10.1103/PhysRevLett.107.123601} {\bibfield
		{journal} {\bibinfo  {journal} {Phys.\ Rev.\ Lett.}\ }\textbf {\bibinfo
			{volume} {107}},\ \bibinfo {pages} {123601} (\bibinfo {year}
		{2011})}\BibitemShut {NoStop}%
	\bibitem [{\citenamefont {Wieczorek}\ \emph {et~al.}(2015)\citenamefont
		{Wieczorek}, \citenamefont {Hofer}, \citenamefont {Hoelscher-Obermaier},
		\citenamefont {Riedinger}, \citenamefont {Hammerer},\ and\ \citenamefont
		{Aspelmeyer}}]{Wieczorek2015}%
	\BibitemOpen
	\bibfield  {author} {\bibinfo {author} {\bibfnamefont {W.}~\bibnamefont
			{Wieczorek}}, \bibinfo {author} {\bibfnamefont {S.~G.}\ \bibnamefont
			{Hofer}}, \bibinfo {author} {\bibfnamefont {J.}~\bibnamefont
			{Hoelscher-Obermaier}}, \bibinfo {author} {\bibfnamefont {R.}~\bibnamefont
			{Riedinger}}, \bibinfo {author} {\bibfnamefont {K.}~\bibnamefont {Hammerer}},
		\ and\ \bibinfo {author} {\bibfnamefont {M.}~\bibnamefont {Aspelmeyer}},\
	}\href {\doibase 10.1103/PhysRevLett.114.223601} {\bibfield  {journal}
		{\bibinfo  {journal} {Phys.\ Rev.\ Lett.}\ }\textbf {\bibinfo {volume}
			{114}},\ \bibinfo {pages} {223601} (\bibinfo {year} {2015})}\BibitemShut
	{NoStop}%
	\bibitem [{\citenamefont {Asano}\ \emph {et~al.}(2017)\citenamefont {Asano},
		\citenamefont {Ochi}, \citenamefont {Takahashi}, \citenamefont {Kishimoto},\
		and\ \citenamefont {Noda}}]{Asano2017}%
	\BibitemOpen
	\bibfield  {author} {\bibinfo {author} {\bibfnamefont {T.}~\bibnamefont
			{Asano}}, \bibinfo {author} {\bibfnamefont {Y.}~\bibnamefont {Ochi}},
		\bibinfo {author} {\bibfnamefont {Y.}~\bibnamefont {Takahashi}}, \bibinfo
		{author} {\bibfnamefont {K.}~\bibnamefont {Kishimoto}}, \ and\ \bibinfo
		{author} {\bibfnamefont {S.}~\bibnamefont {Noda}},\ }\href {\doibase
		10.1364/OE.25.001769} {\bibfield  {journal} {\bibinfo  {journal} {Opt.\
				Express}\ }\textbf {\bibinfo {volume} {25}},\ \bibinfo {pages} {1769}
		(\bibinfo {year} {2017})}\BibitemShut {NoStop}%
	\bibitem [{\citenamefont {Patel}\ \emph {et~al.}(2017)\citenamefont {Patel},
		\citenamefont {Sarabalis}, \citenamefont {Jiang}, \citenamefont {Hill},\ and\
		\citenamefont {Safavi-Naeini}}]{Patel2017}%
	\BibitemOpen
	\bibfield  {author} {\bibinfo {author} {\bibfnamefont {R.~N.}\ \bibnamefont
			{Patel}}, \bibinfo {author} {\bibfnamefont {C.~J.}\ \bibnamefont
			{Sarabalis}}, \bibinfo {author} {\bibfnamefont {W.}~\bibnamefont {Jiang}},
		\bibinfo {author} {\bibfnamefont {J.~T.}\ \bibnamefont {Hill}}, \ and\
		\bibinfo {author} {\bibfnamefont {A.~H.}\ \bibnamefont {Safavi-Naeini}},\
	}\href {\doibase 10.1103/PhysRevApplied.8.041001} {\bibfield  {journal}
		{\bibinfo  {journal} {Phys.\ Rev.\ Applied}\ }\textbf {\bibinfo {volume}
			{8}},\ \bibinfo {pages} {041001} (\bibinfo {year} {2017})}\BibitemShut
	{NoStop}%
	\bibitem [{\citenamefont {Maring}\ \emph {et~al.}(2017)\citenamefont {Maring},
		\citenamefont {Farrera}, \citenamefont {Kutluer}, \citenamefont {Mazzera},
		\citenamefont {Heinze},\ and\ \citenamefont {de~Riedmatten}}]{Maring2017}%
	\BibitemOpen
	\bibfield  {author} {\bibinfo {author} {\bibfnamefont {N.}~\bibnamefont
			{Maring}}, \bibinfo {author} {\bibfnamefont {P.}~\bibnamefont {Farrera}},
		\bibinfo {author} {\bibfnamefont {K.}~\bibnamefont {Kutluer}}, \bibinfo
		{author} {\bibfnamefont {M.}~\bibnamefont {Mazzera}}, \bibinfo {author}
		{\bibfnamefont {G.}~\bibnamefont {Heinze}}, \ and\ \bibinfo {author}
		{\bibfnamefont {H.}~\bibnamefont {de~Riedmatten}},\ }\href {\doibase
		10.1038/nature24468} {\bibfield  {journal} {\bibinfo  {journal} {Nature}\
		}\textbf {\bibinfo {volume} {551}},\ \bibinfo {pages} {485} (\bibinfo {year}
		{2017})}\BibitemShut {NoStop}%
	\bibitem [{\citenamefont {Rueda}\ \emph {et~al.}(2016)\citenamefont {Rueda},
		\citenamefont {Sedlmeir}, \citenamefont {Collodo}, \citenamefont {Vogl},
		\citenamefont {Stiller}, \citenamefont {Schunk}, \citenamefont {Strekalov},
		\citenamefont {Marquardt}, \citenamefont {Fink}, \citenamefont {Painter},
		\citenamefont {Leuchs},\ and\ \citenamefont {Schwefel}}]{Rueda2016}%
	\BibitemOpen
	\bibfield  {author} {\bibinfo {author} {\bibfnamefont {A.}~\bibnamefont
			{Rueda}}, \bibinfo {author} {\bibfnamefont {F.}~\bibnamefont {Sedlmeir}},
		\bibinfo {author} {\bibfnamefont {M.~C.}\ \bibnamefont {Collodo}}, \bibinfo
		{author} {\bibfnamefont {U.}~\bibnamefont {Vogl}}, \bibinfo {author}
		{\bibfnamefont {B.}~\bibnamefont {Stiller}}, \bibinfo {author} {\bibfnamefont
			{G.}~\bibnamefont {Schunk}}, \bibinfo {author} {\bibfnamefont {D.~V.}\
			\bibnamefont {Strekalov}}, \bibinfo {author} {\bibfnamefont {C.}~\bibnamefont
			{Marquardt}}, \bibinfo {author} {\bibfnamefont {J.~M.}\ \bibnamefont {Fink}},
		\bibinfo {author} {\bibfnamefont {O.}~\bibnamefont {Painter}}, \bibinfo
		{author} {\bibfnamefont {G.}~\bibnamefont {Leuchs}}, \ and\ \bibinfo {author}
		{\bibfnamefont {H.~G.~L.}\ \bibnamefont {Schwefel}},\ }\href {\doibase
		10.1364/OPTICA.3.000597} {\bibfield  {journal} {\bibinfo  {journal} {Optica}\
		}\textbf {\bibinfo {volume} {3}},\ \bibinfo {pages} {597} (\bibinfo {year}
		{2016})}\BibitemShut {NoStop}%
	\bibitem [{\citenamefont {Higginbotham}\ \emph {et~al.}(2017)\citenamefont
		{Higginbotham}, \citenamefont {Burns}, \citenamefont {Urmey}, \citenamefont
		{Peterson}, \citenamefont {Kampel}, \citenamefont {Brubaker}, \citenamefont
		{Smith}, \citenamefont {Lehnert},\ and\ \citenamefont
		{Regal}}]{Higginbotham2017}%
	\BibitemOpen
	\bibfield  {author} {\bibinfo {author} {\bibfnamefont {A.~P.}\ \bibnamefont
			{Higginbotham}}, \bibinfo {author} {\bibfnamefont {P.~S.}\ \bibnamefont
			{Burns}}, \bibinfo {author} {\bibfnamefont {M.~D.}\ \bibnamefont {Urmey}},
		\bibinfo {author} {\bibfnamefont {R.~W.}\ \bibnamefont {Peterson}}, \bibinfo
		{author} {\bibfnamefont {N.~S.}\ \bibnamefont {Kampel}}, \bibinfo {author}
		{\bibfnamefont {B.~M.}\ \bibnamefont {Brubaker}}, \bibinfo {author}
		{\bibfnamefont {G.}~\bibnamefont {Smith}}, \bibinfo {author} {\bibfnamefont
			{K.~W.}\ \bibnamefont {Lehnert}}, \ and\ \bibinfo {author} {\bibfnamefont
			{C.~A.}\ \bibnamefont {Regal}},\ }\href {https://arxiv.org/abs/1712.06535}
	{\bibfield  {journal} {\bibinfo  {journal} {arXiv:1712.06535}\ } (\bibinfo
		{year} {2017})}\BibitemShut {NoStop}%
	\bibitem [{\citenamefont {Fang}\ \emph {et~al.}(2017)\citenamefont {Fang},
		\citenamefont {Luo}, \citenamefont {Metelmann}, \citenamefont {Matheny},
		\citenamefont {Marquardt}, \citenamefont {Clerk},\ and\ \citenamefont
		{Painter}}]{Fang2017}%
	\BibitemOpen
	\bibfield  {author} {\bibinfo {author} {\bibfnamefont {K.}~\bibnamefont
			{Fang}}, \bibinfo {author} {\bibfnamefont {J.}~\bibnamefont {Luo}}, \bibinfo
		{author} {\bibfnamefont {A.}~\bibnamefont {Metelmann}}, \bibinfo {author}
		{\bibfnamefont {M.~H.}\ \bibnamefont {Matheny}}, \bibinfo {author}
		{\bibfnamefont {F.}~\bibnamefont {Marquardt}}, \bibinfo {author}
		{\bibfnamefont {A.~A.}\ \bibnamefont {Clerk}}, \ and\ \bibinfo {author}
		{\bibfnamefont {O.}~\bibnamefont {Painter}},\ }\href {\doibase
		10.1038/nphys4009} {\bibfield  {journal} {\bibinfo  {journal} {Nature Phys.}\
		}\textbf {\bibinfo {volume} {13}},\ \bibinfo {pages} {465} (\bibinfo {year}
		{2017})}\BibitemShut {NoStop}%
	\bibitem [{\citenamefont {Cumming}(1957)}]{Cumming1957}%
	\BibitemOpen
	\bibfield  {author} {\bibinfo {author} {\bibfnamefont {R.~C.}\ \bibnamefont
			{Cumming}},\ }\href {\doibase 10.1109/JRPROC.1957.278387} {\bibfield
		{journal} {\bibinfo  {journal} {Proc.\ IRE}\ }\textbf {\bibinfo {volume}
			{45}},\ \bibinfo {pages} {175} (\bibinfo {year} {1957})}\BibitemShut
	{NoStop}%
	\bibitem [{\citenamefont {Wong}\ \emph {et~al.}(1982)\citenamefont {Wong},
		\citenamefont {Rue},\ and\ \citenamefont {Wright}}]{Wong1982}%
	\BibitemOpen
	\bibfield  {author} {\bibinfo {author} {\bibfnamefont {K.~K.}\ \bibnamefont
			{Wong}}, \bibinfo {author} {\bibfnamefont {R.~M. D.~L.}\ \bibnamefont {Rue}},
		\ and\ \bibinfo {author} {\bibfnamefont {S.}~\bibnamefont {Wright}},\ }\href
	{\doibase 10.1364/OL.7.000546} {\bibfield  {journal} {\bibinfo  {journal}
			{Opt.\ Lett.}\ }\textbf {\bibinfo {volume} {7}},\ \bibinfo {pages} {546}
		(\bibinfo {year} {1982})}\BibitemShut {NoStop}%
	\bibitem [{\citenamefont {Horodecki}\ \emph {et~al.}(2009)\citenamefont
		{Horodecki}, \citenamefont {Horodecki}, \citenamefont {Horodecki},\ and\
		\citenamefont {Horodecki}}]{Horodecki2009}%
	\BibitemOpen
	\bibfield  {author} {\bibinfo {author} {\bibfnamefont {R.}~\bibnamefont
			{Horodecki}}, \bibinfo {author} {\bibfnamefont {P.}~\bibnamefont
			{Horodecki}}, \bibinfo {author} {\bibfnamefont {M.}~\bibnamefont
			{Horodecki}}, \ and\ \bibinfo {author} {\bibfnamefont {K.}~\bibnamefont
			{Horodecki}},\ }\href {\doibase 10.1103/RevModPhys.81.865} {\bibfield
		{journal} {\bibinfo  {journal} {Rev.\ Mod.\ Phys.}\ }\textbf {\bibinfo
			{volume} {81}},\ \bibinfo {pages} {865} (\bibinfo {year} {2009})}\BibitemShut
	{NoStop}%
	\bibitem [{\citenamefont {Hill}\ and\ \citenamefont
		{Wootters}(1997)}]{Hill1997}%
	\BibitemOpen
	\bibfield  {author} {\bibinfo {author} {\bibfnamefont {S.}~\bibnamefont
			{Hill}}\ and\ \bibinfo {author} {\bibfnamefont {W.~K.}\ \bibnamefont
			{Wootters}},\ }\href {\doibase 10.1103/PhysRevLett.78.5022} {\bibfield
		{journal} {\bibinfo  {journal} {Phys.\ Rev.\ Lett.}\ }\textbf {\bibinfo
			{volume} {78}},\ \bibinfo {pages} {5022} (\bibinfo {year}
		{1997})}\BibitemShut {NoStop}%
	\bibitem [{\citenamefont {Hofer}\ \emph {et~al.}(2011)\citenamefont {Hofer},
		\citenamefont {Wieczorek}, \citenamefont {Aspelmeyer},\ and\ \citenamefont
		{Hammerer}}]{Hofer2011}%
	\BibitemOpen
	\bibfield  {author} {\bibinfo {author} {\bibfnamefont {S.~G.}\ \bibnamefont
			{Hofer}}, \bibinfo {author} {\bibfnamefont {W.}~\bibnamefont {Wieczorek}},
		\bibinfo {author} {\bibfnamefont {M.}~\bibnamefont {Aspelmeyer}}, \ and\
		\bibinfo {author} {\bibfnamefont {K.}~\bibnamefont {Hammerer}},\ }\href
	{\doibase 10.1103/PhysRevA.84.052327} {\bibfield  {journal} {\bibinfo
			{journal} {Phys.\ Rev.\ A}\ }\textbf {\bibinfo {volume} {84}},\ \bibinfo
		{pages} {52327} (\bibinfo {year} {2011})}\BibitemShut {NoStop}%
	\bibitem [{\citenamefont {Kuzmich}\ \emph {et~al.}(2003)\citenamefont
		{Kuzmich}, \citenamefont {Bowen}, \citenamefont {Boozer}, \citenamefont
		{Boca}, \citenamefont {Chou}, \citenamefont {Duan},\ and\ \citenamefont
		{Kimble}}]{Kuzmich2003}%
	\BibitemOpen
	\bibfield  {author} {\bibinfo {author} {\bibfnamefont {A.}~\bibnamefont
			{Kuzmich}}, \bibinfo {author} {\bibfnamefont {W.~P.}\ \bibnamefont {Bowen}},
		\bibinfo {author} {\bibfnamefont {A.~D.}\ \bibnamefont {Boozer}}, \bibinfo
		{author} {\bibfnamefont {A.}~\bibnamefont {Boca}}, \bibinfo {author}
		{\bibfnamefont {C.~W.}\ \bibnamefont {Chou}}, \bibinfo {author}
		{\bibfnamefont {L.-M.}\ \bibnamefont {Duan}}, \ and\ \bibinfo {author}
		{\bibfnamefont {H.~J.}\ \bibnamefont {Kimble}},\ }\href {\doibase
		10.1038/nature01714} {\bibfield  {journal} {\bibinfo  {journal} {Nature}\
		}\textbf {\bibinfo {volume} {423}},\ \bibinfo {pages} {731} (\bibinfo {year}
		{2003})}\BibitemShut {NoStop}%
	\bibitem [{\citenamefont {Lee}(2012)}]{LeePhD}%
	\BibitemOpen
	\bibfield  {author} {\bibinfo {author} {\bibfnamefont {K.~C.}\ \bibnamefont
			{Lee}},\ }\emph {\bibinfo {title} {Generation of room-temperature
			entanglement in diamond with broadband pulses}},\ \href
	{https://www2.physics.ox.ac.uk/sites/default/files/2013-11-08/kc_lee_thesis_76347.pdf}
	{Ph.D. thesis},\ \bibinfo  {school} {University of Oxford} (\bibinfo {year}
	{2012})\BibitemShut {NoStop}%
\end{thebibliography}
\end{document}